\documentclass[fleqn,usenatbib]{mnras}

\usepackage{newtxtext,newtxmath}
\usepackage[T1]{fontenc}

\DeclareRobustCommand{\VAN}[3]{#2}
\let\VANthebibliography\thebibliography
\def\thebibliography{\DeclareRobustCommand{\VAN}[3]{##3}\VANthebibliography}

\usepackage{graphicx}	% Including figure files
\usepackage{amsmath}	% Advanced maths commands

\usepackage{amssymb}	% Extra maths symbols

\title[Anisotropies of COB and CNIRB from the CSST]{Anisotropies of Cosmic Optical and Near-IR Background from the China Space Station Telescope (CSST)}

\author[Y. Cao et al.]{
Ye Cao$^{1,2}$, 
Yan Gong$^{1}$\thanks{E-mail: gongyan@bao.ac.cn}, 
Dezi Liu$^{3}$, 
Asantha Cooray$^{4}$, 
Chang Feng$^{5,6}$
and Xuelei Chen$^{7,2,8}$
\\
$^{1}$ Key Laboratory of Space Astronomy and Technology, National Astronomical Observatories, Chinese Academy of Sciences, Beijing 100101, China.\\
$^{2}$ School of Astronomy and Space Sciences, University of Chinese Academy of Sciences, Beijing 100049, China\\
$^{3}$ South-Western Institute for Astronomy Research, Yunnan University, Kunming 650500, China\\
$^{4}$ Department of Physics and Astronomy, University of California, Irvine, CA 92697, USA\\
$^{5}$ CAS Key Laboratory for Researches in Galaxies and Cosmology, Department of Astronomy, University of Science and Technology of China,\\  Chinese Academy of Sciences, Hefei, Anhui 230026, China\\
$^{6}$ School of Astronomy and Space Science, University of Science and Technology of China, Hefei, 230026, China\\
$^{7}$ Key Laboratory of Computational Astrophysics, National Astronomical Observatories, Chinese Academy of Sciences, Beijing 100101, China\\
$^{8}$ Center for High Energy Physics, Peking University, Beijing 100871, China
}

\date{Accepted XXX. Received YYY; in original form ZZZ}

\pubyear{2021}

\begin{document}
\label{firstpage}
\pagerange{\pageref{firstpage}--\pageref{lastpage}}
\maketitle

\begin{abstract}
Anisotropies of the cosmic optical background (COB) and cosmic near-IR background (CNIRB) are capable of addressing some of the key questions in cosmology and astrophysics. In this work, we measure and analyze the angular power spectra of the simulated COB and CNIRB in the ultra-deep field of the China Space Station Telescope (CSST-UDF). The CSST-UDF covers about 9 square degrees, with magnitude limits $\sim$28.3, 28.2, 27.6, 26.7 AB mag for point sources with 5$\sigma$ detection in the $r$ (0.620$\rm \mu m$), $i$ (0.760$\rm \mu m$), $z$ (0.915$\rm \mu m$), and $y$ (0.965$\rm \mu m$) bands, respectively. According to the design parameters and scanning pattern of the CSST, we generate mock data, merge images and mask the bright sources in the four bands. We obtain four angular power spectra from $\ell=200$ to 2,000,000 (from arcsecond to degree), and fit them with a multi-component model including intrahalo light (IHL) using the Markov chain Monte Carlo (MCMC) method. We find that, the signal to noise ratio (SNR) of the IHL is larger than 8 over the range of angular scales that is useful for astrophysical studies ($\ell\sim$10,000-400,000). Comparing to previous works, the constraints on the model parameters are improved by factors of 3$\sim$4 in this study, which indicates that the CSST-UDF survey can be a powerful probe on the cosmic optical and near-IR backgrounds.
\end{abstract}


%%%%%%%%%%%%%%%%%%%%%%%%%%%%%%%%%%%%%%%%%%%%%%%%%%

%%%%%%%%%%%%%%%%% BODY OF PAPER %%%%%%%%%%%%%%%%%%

\section{Introduction}

In the late Dark Ages, dark matter and clustering of gas distribution formed in the universe as primary density perturbation grow under the gravity. As the gas collapsed and the dark matter halos merged, the first galaxies in the universe began to form and evolve. Then the neutral hydrogen around the galaxy began to be ionized by the high-energy photons from the massive stars, forming ionization regions around the surrounding galaxies, and the universe began to reionize during the epoch of reionization (EoR). The luminosity density of thtese ultraviolet photons (rest wavelength $\sim$150$\rm \mu m$) is an important observational goal for investigating the formation and early evolution. As proposed by previous works, the luminosity density $\rho_{\rm UV}$ can be obtained by measuring the extragalactic background light \citep{Mitchell-Wynne15}. 

Since the cosmic optical background (COB) and cosmic near-IR background (CNIRB) are both the accumulation of all the emission throughout the universe history, the COB and CNIRB fluctuations arise from the emission of several sources, e.g. galaxies from low and high redshifts, the intrahalo light (IHL), the diffused Galactic light (DGL), etc. According to the Planck's measurements of reionization optical depth \citep{Planck16XIII}, the extragalactic background light (EBL) has a peak between optical and near-IR wavelengths. Since Lyman break during the EoR is redshifted to about 0.8$\rm \mu m$ today and the components of blueward of Lyman break do not contribute to the results, the intensity fluctuations of COB and CNIRB with wavelengths around 1$\rm \mu m$ provide the best mechanism. \cite{Mitchell-Wynne15} proposed that the emission from the EoR could be measured indirectly by separating other components. Therefore, accurate measurement of the emission from low-$z$ galaxies, DGL and IHL can be helpful to separate the signal from the EoR.

The intensity fluctuations of the COB and CNIRB have been studied in the literature. \cite{Dube77,Dube79} first measured the optical background by subtracting all foreground components from the observed brightness of the night sky. Then \cite{Mattila76} attempted to to shield the foreground components by using Galactic dark clouds. But these attempts are not very successful due to a lack of understanding of foreground radiations, \cite{Bernstein02a,Bernstein02b,Bernstein02c} solved this problem by coordinating data from Hubble Space Telescope (HST) and a ground-based telescope at Las Campanas Observatory, and \cite{Bernstein05} and \cite{Bernstein07} made a series of corrections to the above results. Then \cite{Matsuoka11} presented a new constraints on the optical background by analyzing the Pioneer 10/11 Imaging Photopolarimeter data. The recent detection of the COB comes from the New Horizons Long-range Reconnaissance Imager (LORRI) data given by \cite{Lauer21}. In recent years, a series of more sensitive experimental measures have made major contribution to quantifying the CNIRB anisotropies, such as those of the HST, $Spitzer$ Space Telescope and the Cosmic Infrared Background Experiment (CIBER)\citep{Grogin11,Koekemoer11,Ashby09,Zemcov13}. These data strongly restrict the CNIRB models by enhanced resolution, sensitivity, frequency coverage and detection area. These observations have higher resolution and sensitivity, wider frequency coverage and larger detection area, which provide strong constraints on the CNIRB model.

In this work, we investigate the measurements of COB and CNIRB intensity fluctuations for the China Space Station Telescope (CSST). The CSST is a 2-meter multi-band space telescope planned for the China Manned Space Program, it is planned to be launched around 2024, and it will operate in the same orbit as the China Manned Space Station. The major science project of CSST includes both large-scale multi-color imaging and slitless spectroscopic observations, it shall have excellent performance of large field of view ($\sim$1 ${\rm deg}^2$), high-spatial resolution (80\% energy concentration region is $\sim$0.15 arcsec), faint magnitude limits, and wide wavelength coverage from 0.25 to 1.1$\rm \mu m$ \citep{Zhan11,Cao18,Gong19}. According to the above analysis, the wavelength coverage range of CSST includes the wavelength of the signal form the EoR. CSST will explore some important cosmological and astronomical objectives, such as galaxy, galaxy clusters, active galactic nuclei, weak gravitational lensing and background intensity fluctuations, etc. Therefore, the CSST is expected to be a powerful survey for investigating the fundamental cosmological and astronomical questions such as gravity, dark matter and dark energy, the cosmic large scale structure, galaxy formation and evolution, the formation of supermassive black hole.

In order to investigate the observation ability of the CSST on the background intensity fluctuations in optical and near-infrared wavelengths and the limitations on the parameters of the cosmological models, we use simulations to obtain the mock background images and analyze their power spectra. In this work, we make use of a nine-square-degree ultra-deep field of the China Space Station Telescope (CSST-UDF) that will be scanned 60 times with a total exposure time of 15000 seconds. The magnitude limits are 28.3, 28.2, 27.6, 26.7 for point sources with 5$\sigma$ detection for the $r$, $i$, $z$ and $y$ bands, respectively. We set that different scans in the same pointing cover the same area, and the width of the overlapping of the adjacent tiles is about $10''$. We use the self-calibration technique \citep{Fixsen00} to merge the tiles as mock background maps, then remove the contamination from bright stars and galaxies with a $3\sigma$ flux cut. Following \cite{Cooray12} and \cite{Zemcov14}, the angular power spectra are estimated by the standard Fourier transform, and they are corrected by the beam function, the transfor function and the mode coupling matrix. Here we obtain four auto-power spectra of background intensity fluctuations at the multipoles from 200 to 2,000,000, and then fit the measured power spectra using the theoretical model and perform a Markov chain Monte Carlo (MCMC) analysis to constrain model parameters of the multi-component model.

This paper is organized as follows: in Section \S\ref{model}, we describe the components of the background intensity fluctuations and their theoretical models. In Section \S\ref{data}, we introduce the instrument parameters of the CSST and the method used to create the mock data. In Section \S\ref{spectrum}, we describe the final power spectrum measurements and corrections. In Section \S\ref{spec_result}, we present the measured power spectra and use a multi-component model to fit them by MontePython code \citep{Audren13}. We finally summarize the results in Section \S\ref{conclusions}. In this work, we adopt the standard $\Lambda \mathrm{CDM}$ cosmological model with parameter values derived from \cite{Planck18VI}, which gives $h$ = 0.674, $\Omega_{\mathrm{m}}$ = 0.315, $\Omega_{\mathrm{b}} h^2$ = 0.0224, $\sigma_{8}$ = 0.811, and $n_s$ = 0.965.

\section{theoretical model}
\label{model}

We adopt a multi-component model, which consists of four main components: IHL, DGL, shot noise and high-redshift galaxies. The IHL luminosity as a function of halo mass $M$ and redshift $z$, can be expressed in the following form at rest-frame wavelength $\lambda$ \citep{Cooray12}
\begin{equation}
L_{\lambda}^{\mathrm{IHL}}(M,z)=f_{\mathrm{IHL}}(M) L(M,z=0)(1+z)^{\alpha} f_{\lambda}(\lambda)
\label{eq:L_IHL},
\end{equation}
where $f_{\lambda}$ is the rest-frame spectral energy distribution (SED) used to describe IHL, which is normalized to be 1 at 2.2$\rm \mu m$. And the IHL SED is from models related to old stellar populations \citep{Krick07}. $f_{\rm IHL}$ is the fraction of total halo luminosity, which can be expressed as
\begin{equation}
f_{\mathrm{IHL}}(M)=A_{\mathrm{IHL}}\left(\frac{M}{M_{0}}\right)^{\beta},
\end{equation}
where the model parameters $A_{\rm IHL}$ and $\beta$ are the amplitude factor and mass power index, respectively. Here we normalize to $M_{0} = 10^{12} M_{\odot}$ \citep{Zemcov14}. In Equation.\ref{eq:L_IHL}, $L(M,z=0) = L_0(M,z=0)/\lambda_0$ is the total halo luminosity at $z = 0$, here $\lambda_0 = 2.2{\rm \mu m}$. $L_0(M,z=0)$ is measured at 2.2$\rm \mu m$ as a function of total halo luminosity and halo mass \citep{Lin04}, it can be written as
\begin{equation}
L_{0}(M, z=0)=5.64 \times 10^{12} h_{70}^{-2}\left(\frac{M}{2.7 \cdot 10^{14} h_{70}^{-1} M_{\odot}}\right)^{0.72} L_{\odot},
\end{equation}
where, $H_{0}=70 h_{70} \mathrm{~km} \mathrm{~s}^{-1} \mathrm{Mpc}^{-1}$ is Hubble constant today
$L_{\odot}$ is the total luminosity of the sun. 

Following the halo model, the power spectrum of the IHL component at frequencies $\nu$ is
\begin{equation}
C_{\ell,\nu}^{\mathrm{IHL}}=C_{\ell,\nu}^{\rm 1h}+C_{\ell,\nu}^{\rm 2h}.
\end{equation}
The 1-halo term related to small-scale fluctuations within individual halos, which can be expressed by
\begin{equation}
\begin{split}
C_{\ell,\nu}^{\rm 1h}&= \frac{1}{(4 \pi)^{2}} \int dz\left(\frac{d \chi}{d z}\right)\left(\frac{a}{\chi}\right)^{2} \\
 \times & \int dM \frac{dn(M, z)}{dM} \left[u_{\mathrm{IHL}}(k \mid M) L_{\lambda}^{\rm IHL}(M, z)\right]^2.
\end{split}
\end{equation}
The 2-halo term related to the large-scale matter fluctuations can be calculated as
\begin{equation}
\begin{split}
C_{\ell,\nu}^{\rm 2h}&= \frac{1}{(4 \pi)^{2}} \int d z\left(\frac{d \chi}{d z}\right)\left(\frac{a}{\chi}\right)^{2} P(k=\ell / \chi(z), z) \\
\times &\left[\int dM \frac{dn(M, z)}{dM} u_{\mathrm{IHL}}(k\mid M) b_{h}(M, z) L^{\mathrm{IHL}}_{\lambda}(M, z)\right]^{2}.
\end{split}
\end{equation}
Here $\chi(z)$ is the comoving distance along the line of sight, $P(k, z)$ is the matter power spectrum at redshift $z$. It is found that components at redshift $z \textgreater 6$ have a very weak effect on the integration result, so we fix the maximum redshift $z_{\rm max} = 6$ \citep{Cooray12}. Upper and lower limits of mass integration determine the relative amplitude of the 1-halo and 2-halo terms, and we set the $M_{\rm min}$ and $M_{\rm max}$ to $10^9$ and $10^{13}M_{\odot}/h^{-1}$, respectively. $b_{h}(M, z)$ is the linear bias, which is only accurate on large scales. $u_{\mathrm{IHL}}(k \mid M)$ is the Fourier transform of the dark matter distribution of a halo with mass $M$ and redshift $z$. 

For the NFW profile, the halo concentration parameter $c_{\mathrm{vir}}$ can be obtained by numerical simulations \citep{Bullock01,Cooray12}, this becomes
\begin{equation}
c_{\mathrm{vir}}(M, z)=\frac{9}{1+z}\left(\frac{M}{M_{*}}\right)^{-0.13},
\end{equation}
where $M_{*}$ is characteristic mass scale at which $\nu\equiv \delta_c^2/\sigma^2(M_*)  =1$, here $\delta_c$ is the critical density contrast required for spherical collapse, $\sigma^2(M)$ is the variance in the initial density fluctuation field on the scale of an object with mass $M$.

DGL component from dust-scattered light of the interstellar radiation field \citep{Zemcov14}, and it is proportional to the dust size and density, so the angular power spectrum related to the interstellar dust emission. The DGL is dominates at the larger scales, and  $C_{\ell}\propto \ell^{-3}$. We set a model parameter $A_{\rm DGL}$ as the amplitude factor at a uniform interstellar radiation field, 
\begin{equation}
C_{\ell}^{\rm{DGL}}=A_{\rm DGL}\ell^{-3}.
\end{equation}

The central wavelength of the $y$ band is 0.965 $\rm{\mu m}$, so the extragalactic background contain not only IHL but also some EoR signal from the high-$z$ galaxies at CSST-UDF $y$ band. For the other CSST bands, the EoR signal can be safely ignored since they locate at shorter wavelenths. The component form the galaxies during reionization is based on an analytic model discussed by \cite{Fernandez12} and \cite{Cooray12apj}. The angular power spectrum of the emission from the high-$z$ galaxies at observed frequencies $\nu$ is
\begin{equation}
C_{\ell,\nu}^{\rm high-z}=\int_{z_{\min }}^{z_{\max }} d z\left(\frac{d \chi}{d z}\right)\left[\frac{a}{\chi} \bar{j}_{\nu}(z)\right]^{2} P_{\mathrm{gg}}(k, z),
\end{equation}
where $P_{\mathrm{gg}}(k, z)$ is the galaxy power spectrum at wavenumber $k$ and redshift $z$, and we set $z_{\rm min}=6$ and $z_{\rm max}=30$. $\bar{j}_{\nu}(z)$ is the mean emissivity produced by the PopII and early PopIII stars, which can be written as
\begin{equation}
\bar{j}_{\nu}(z)=\frac{1}{4\pi}\left[f_{p}l_{\nu}^{\rm II}\left\langle\tau_{*}^{\rm II}\right\rangle+\left(1-f_{p}\right)l_{\nu}^{\rm III}\left\langle\tau_{*}^{\rm III}\right\rangle\right] \psi(z),
\label{eq:j_u}
\end{equation}
where $l_{\nu}$ and $\left\langle\tau_{*}\right\rangle$ are the luminosity mass density and the average stellar lifetime, respectively. The superscript ${\rm II}$ and ${\rm III}$ are shorthand for PopII and early PopIII stars, and the relative fraction of the Pop II and Pop III stars at redshift $z$ is
\begin{equation}
f_{p}(z)=\frac{1}{2}\left[1+\operatorname{erf}\left(\frac{z-10}{\sigma_{p}}\right)\right],
\end{equation}
where $\sigma_{p}$ = 0.5 is the population transition width. In Equation (\ref{eq:j_u}), $\psi(z)$ is the comoving star formation rate density, which is given by
\begin{equation}
\psi(z)=f_{*} \frac{\Omega_{\mathrm{b}}}{\Omega_{\mathrm{m}}} \frac{d}{d t} \int_{M_{\min }}^{\infty} d M\ M \frac{d n}{d M}(M, z).
\end{equation}

For simplicity, we fix the model parameters of high-redshift galaxies, and just fit the amplitude scales $A_{\rm high-z}$. We set the star formation efficiency $f_{*}=0.03$, the threshold mass $M_{\rm min}=5\times10^7M_{\odot}$ and the escape fraction of the Lyman-$\alpha$ photons $f_{\rm esc}=0.2$ \citep{Mitchell-Wynne15}. The optical depth to reionization of this model is 0.07, which is consistent with the $\it Planck$ result \citep{Planck16XIII}.

The measurement of power spectrum is inevitably affected by shot noise, which dominates the optical and infrared background fluctuation at the small scales. The power spectrum of shot noise is usually approximated as a constant, so that the final angular power spectrum is
\begin{equation}
\label{eq:Cl}
C_{\ell}= \begin{cases}C_{\ell}^{\mathrm{IHL}}+C_{\ell}^{\mathrm{DGL}}+C_{\ell}^{\text {high-z}}+N_{\text {shot }} & \text {y band,} \\ C_{\ell}^{\mathrm{IHL}}+C_{\ell}^{\mathrm{DGL}}+N_{\text {shot }} & \text {others.}\end{cases}
\end{equation}

In addition to the above components, the residue galaxy clustering signal from the low redshift source is also included in the background. We can calculate the fraction of residual low-$z$ galaxies by using the model given by \citep{Helgason12}, but we find that it has much less influence on the shape of the final background power spectrum than the other components. Hence we ignore this component in this work \citep[also see e.g.][]{Mitchell-Wynne15}.

\section{Mock data}
\label{data}

In this section, we introduce the method used to create the mock data of the CSST-UDF. There are two steps to obtain the observed sky map: constructing the mock galaxy catalog and generating the simulated image. The mock galaxy catalog contain the information about the multicolor photometry, shapes and positions of galaxies and stars. After obtaining the mock galaxy catalog, we create a map of background intensity fluctuations map, and add stars and galaxies from the mock catalog to the sky map. Then we add the zodiacal light, earthshine and instrumental noise to the mock data.

\subsection{Mock galaxy catalog}
\label{subsec:catalog}

In this work, the mock catalogs we generate consist mainly of stars and galaxies. Since we are focusing on the background, for simplicity, we assume that the galaxies and stars are randomly distributed on the sky map, and the information of clustering is not considered. 

For stars, since they are point sources, we can set the surface brightness distribution as a $\delta$ function. We follow \cite{Robin03} in studying structure and evolution of the Milky Way. We assume that the magnitude distribution of stars in the $i$ band satisfies the Besancon structure model. According to previous studies, the magnitude distributions of stars are almost identical between the Galactic latitude $50$ degrees and $90$ degrees \citep[e.g.][]{Hoekstra17}. Therefore, we adopt  their results to model the magnitude distribution of stars at high Galactic latitude.

We generate the stellar components in the sky map based on the template library obtained by \cite{Polletta07}, which are obtained by studying the stellar spectrum. The templates of stars spectral energy distributions (SEDs) includes 131 stellar templates (from type O to type M). Following \cite{Cao18} and \cite{Zhou21}, we calculate the mocked flux in each band by convolving the SED with the CSST filter response. 

The characteristics of galaxies are much more complex than those of stars. Here, we assume that the surface brightness distribution of the galaxy contains two components: the exponential disc distribution and the de Vaucouleurs bulge distribution \citep{deVaucouleurs48}. We assume that the relative intensity between the two components in a galaxy is determined by the bulge-to-total flux ratios (B/T) which follows a truncated Gaussian distribution $N(0.0, 0.1^2)$ in the range of $[0.0, 1.0]$, and around 10 percent of galaxies are set to be bulge-dominated with B/T=1.0 \citep{Miller13}. 

The magnitude distribution of our simulated galaxy samples in the $i$ band is based on the second-order polynomial fitting of the data in the F775W band at HST-GOODS-NORTH, HST-GOODS-SOUTH and HST-UVUDF field. we adopt the redshift distribution function studied by \cite{Raichoor14}, which is derived from spectral data such as VVDS and SDSS, and we assume that there is a correlation between the redshift and the magnitude. The the galaxy SED templates are based on the $BPz$ template library \citep{Benitez00}, which includes eight templates: one elliptical galaxy SED, two spiral galaxies, two starburst galaxy, one irregular galaxy and two star-forming galaxy. 

Five galaxy intrinsic extinction laws are considered, which are derived from the studies of the Milky Way (MW) \citep{Allen76,Seaton79}, Large Magellanic Cloud (LMC) \citep{Fitzpatrick86}, Small Magellanic Cloud (SMC) \citep{Prevot84,Bouchet85}, and starburst galaxy (SBG) \citep{Calzetti00}. The extinction of intergalactic medium (IGM) has also been considered for all types of galaxies, and the IGM extinction model is based on the results of \cite{Madau95}. For mock photometric data of the galaxies, they are obtained by convolving the redshifted galaxy SED with the CSST filter response \citep{Cao18}. 

Both the galaxy size and ellipticity distributions are given in the CFHTLenS simulations \citep{Miller13}. The relationship between the $i$-band magnitude and median major-axis scalelength for exponential component  is obtained by fitting the disc-dominated galaxies in HST-Groth field. For the bulge component, its half-light radius is 0.6 times the corresponding exponential component. The ellipticity distribution of disc-dominated galaxies is formulated by fitting the disc-dominated galaxies selected from SDSS DR7 \citep{Abazajian09}, while the ellipticity distribution of bulge-dominated galaxies is constructed by fitting the bulge-dominated galaxy sample \citep{Simard02}.  Finally, the orientations of the galaxies are randomly assigned, following a uniform distribution on the interval $[-\pi/2, +\pi/2]$.
The final catalog includes the magnitude of each band, the redshift, the position of the sky, the parameters of surface brightness, and the galaxy shape parameters.

\subsection{Mock sky map}
\label{subsec:map}

We produce mock sky realizations from the theoretical angular power spectra with the instrumental noise and sky background, and add the sources from the catalogs obtain in the previous subsection. Then we obtain the simulated images of each exposure according to the scanning pattern of CSST, and merge them into a final sky map.

The background angular power spectrum can be obtained by Equation (\ref{eq:Cl}), and we set $\log_{10}(A_{\rm IHL}) = -3.23$, $\log_{10}(A_{\rm high-z}) = 1.19$, $\alpha = 0.094$, and $\beta = 1.0$. The DGL amplitude and the shot noise are assumed to be $A_{\rm DGL}= \lbrace 1750,\ 2710,\ 2910,\ 3050 \rbrace$ and $N_{\rm shot}=\lbrace 3.12,\ 4.29,\ 11.9,\ 18.4 \rbrace \times 10^{-12}$ for the $r$, $i$, $z$ and $y$ band, respectively \citep{Cooray12,Mitchell-Wynne15}. We produce mock sky realization from the theoretical angular power spectrum. The mock map cover more than $3\times 3$ square degrees of the sky.

The sky image simulations are based on the GALSIM image simulation software \citep{Rowe15}. The size of a single CSST charge coupled device (CCD) detectors is ${\rm 9k \times 9k}$, and the pixel scale is $0.074''$, so we can get about a $11' \times 11'$ tile of sky with each exposure at each band. Then we add the sources to the background within the area of tile. The relationship between apparent magnitude $m$ of the source and the integrated CCD counts $f_{\rm CDD}$ in unit of Analog-Digital Unit (ADU) is
\begin{equation}
m = -2.5\log(f_{\rm CCD})+Zp,
\end{equation}
where $Zp$ is the zero point of the filters used on CSST, and we can define
\begin{equation}
Zp=m_{\rm Lim}+2.5\log \left(n \sigma t \sqrt{N_{\rm pix}}\right).
\end{equation}
Here the limiting magnitude $m_{\rm Lim}$ is 28.3, 28.2, 27.6 and 26.7 (which are $\{1.36,\ 0.98,\ 1.21,\ 2.44\}\times10^{-14} {\rm nW/m^2/}${\AA} for point sources with 5$\sigma$ detection) for $r$, $i$, $z$ and $y$ band, respectively. $n$ is the signal to noise ratio (SNR) for each source in the catalog based on the CSST instrumental parameters. $\sigma$ is the standard deviation of noise. $N_{\rm pix}$ is the number of pixels covered by the source on the CCD, and $t = 250{\rm s}$ is the exposure time.

For simplicity, we also need to consider the beam of the instrument. We assume that CSST has symmetric 2D-Gaussian beam in each band, and it does not change with the position of the focal plane. We convolve the beam with the background image containing galaxies and stars to obtain a simulated tile without noise. For static point source in the tile, the radius at 80\% of the total energy is 0.135, 0.145, 0.165 and 0.165 arc seconds for $r$, $i$, $z$ and $y$ band, respectively.

\begin{figure}
\centering
\includegraphics[width=1.\columnwidth]{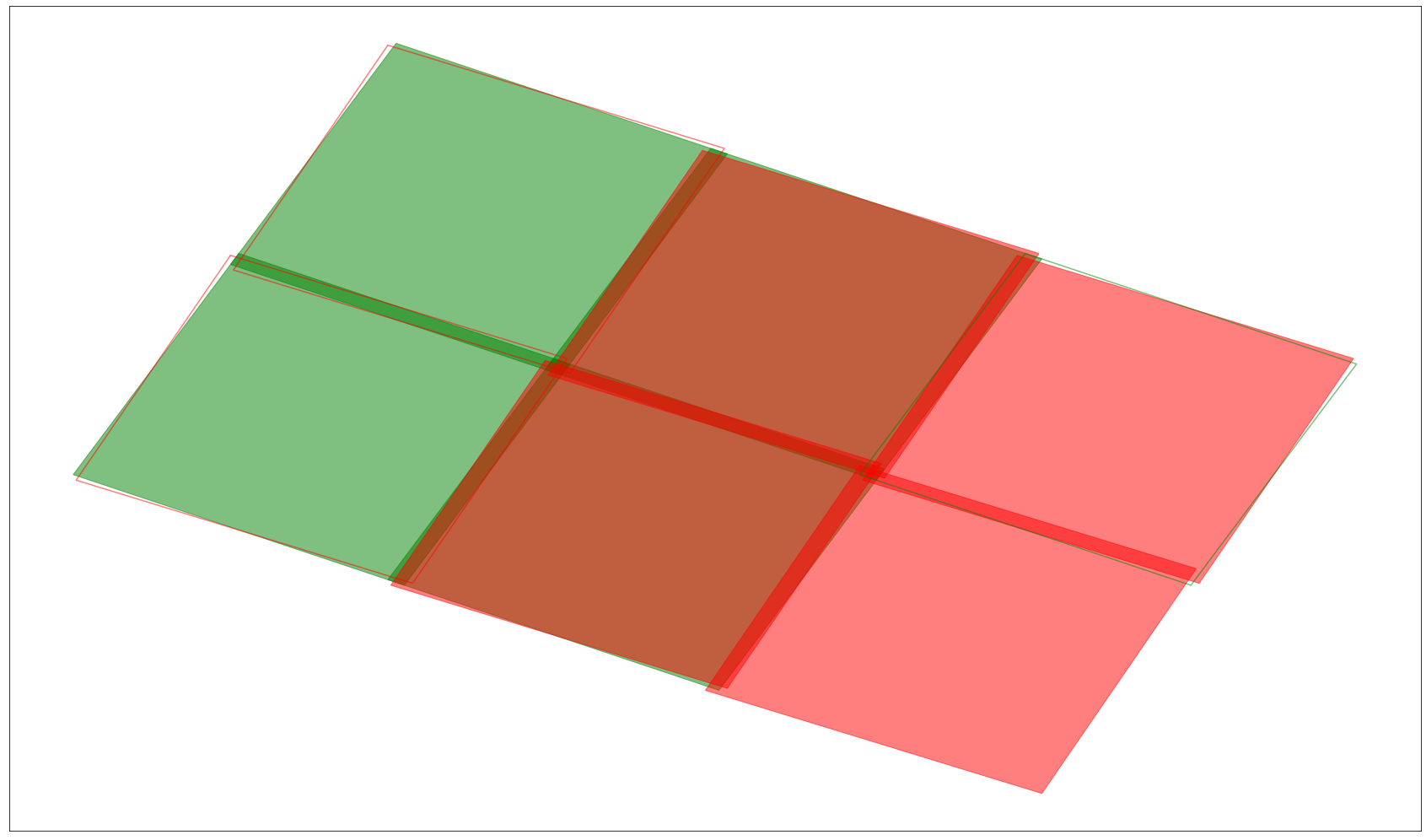}
\caption{ An example of tile pattern. We present 8 tiles from two scans (note that the area in the middle contains two tiles), here each scan consists of 4 tiles (represent by red and green quadrilaterals). We can find that the coverage of two exposures with the same pointing is same ( two fields in the middle ), and the width of the overlapping of two adjacent tiles is about $10''$. }
\label{fig:ccd} 
\end{figure}

\begin{figure}
\centering
\includegraphics[width=1.\columnwidth]{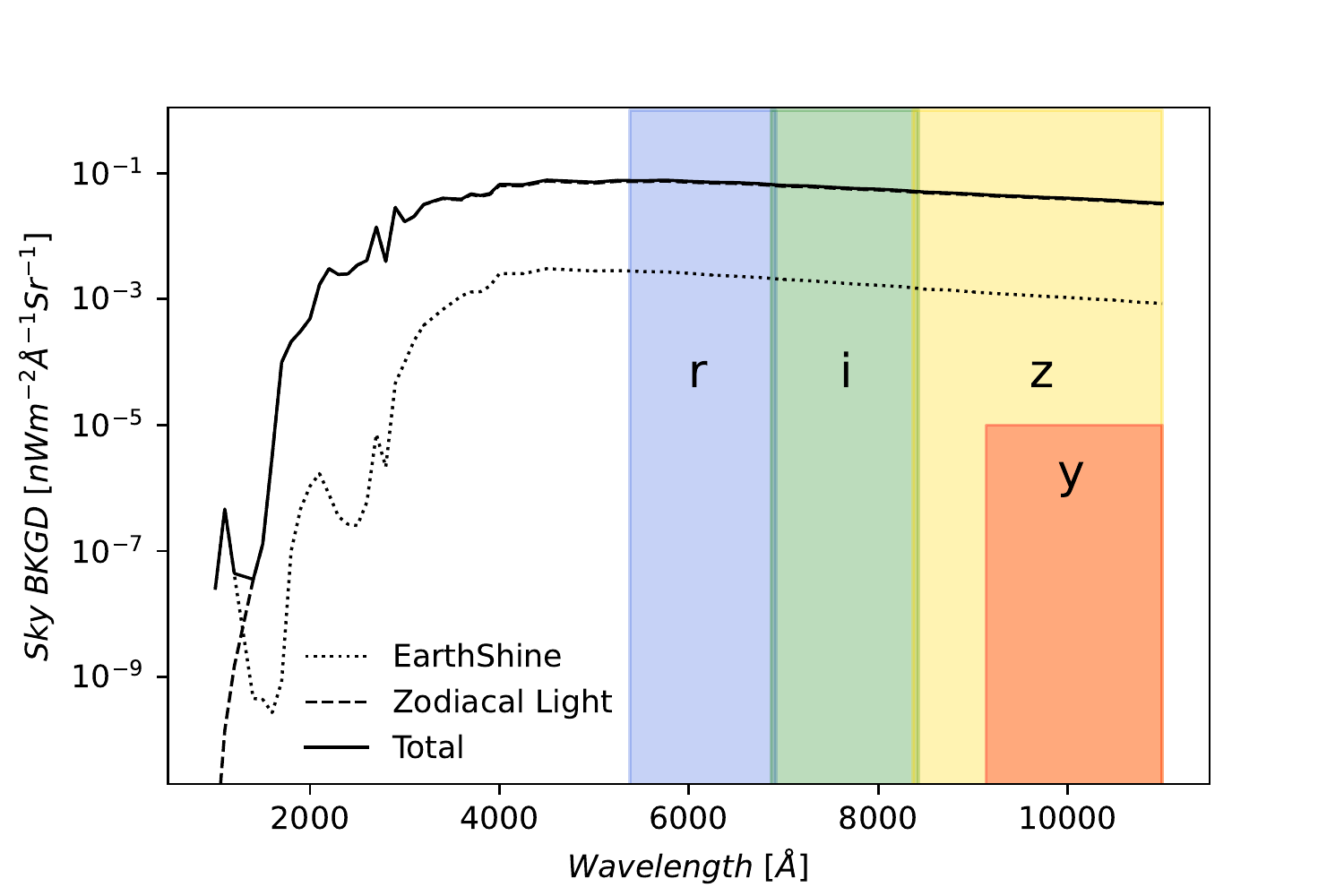}
\caption{ The sky background intensity as a function of wavelength. The dotted, dashed, and solid curves are the earthshine, zodiacal light, and total sky background, respectively. The colored areas indicate the coverages of the four filters.}
\label{fig:skybackground} 
\end{figure}

\begin{figure*}
\centering
\includegraphics[width=1.\columnwidth]{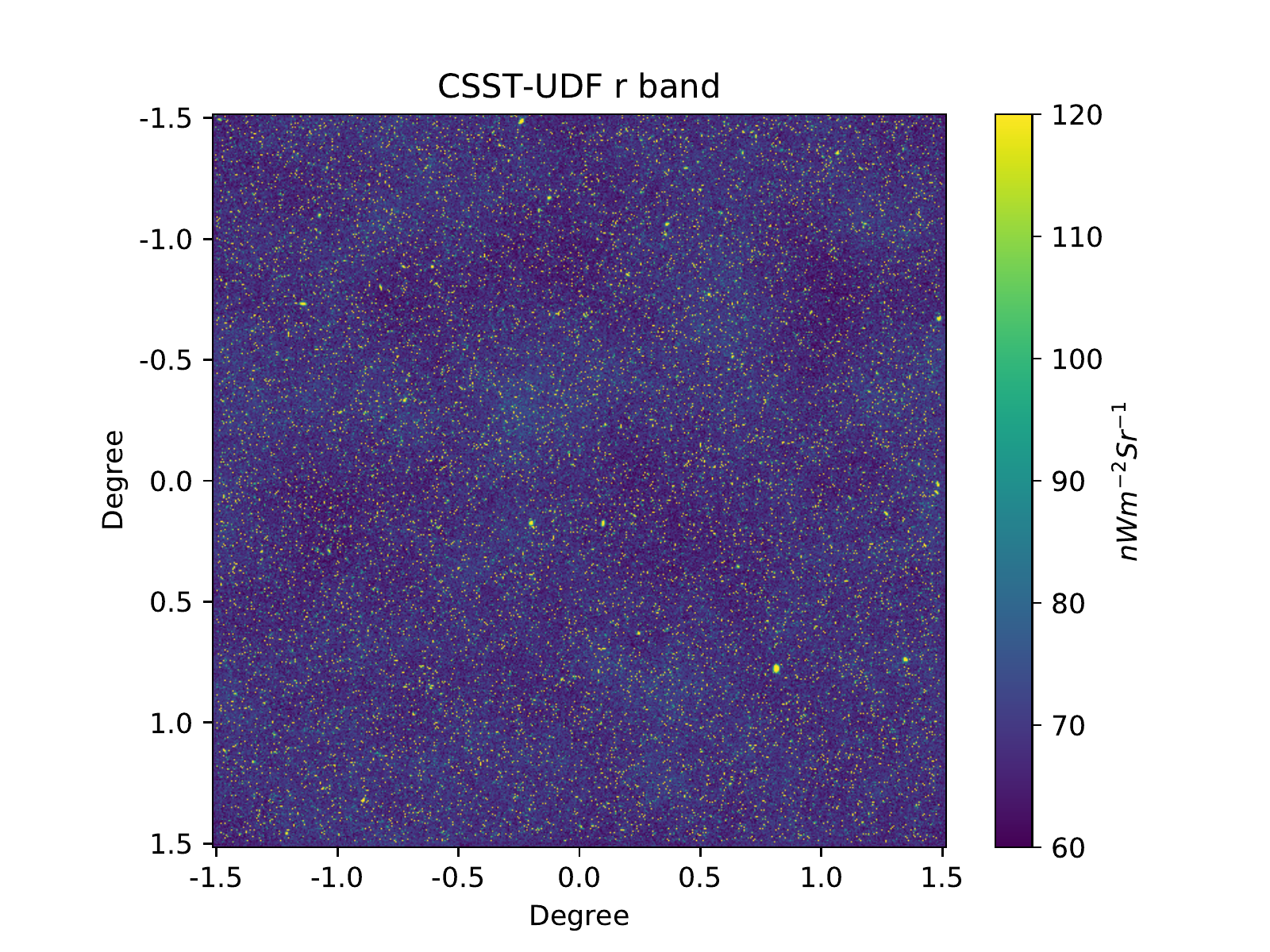}
\includegraphics[width=1.\columnwidth]{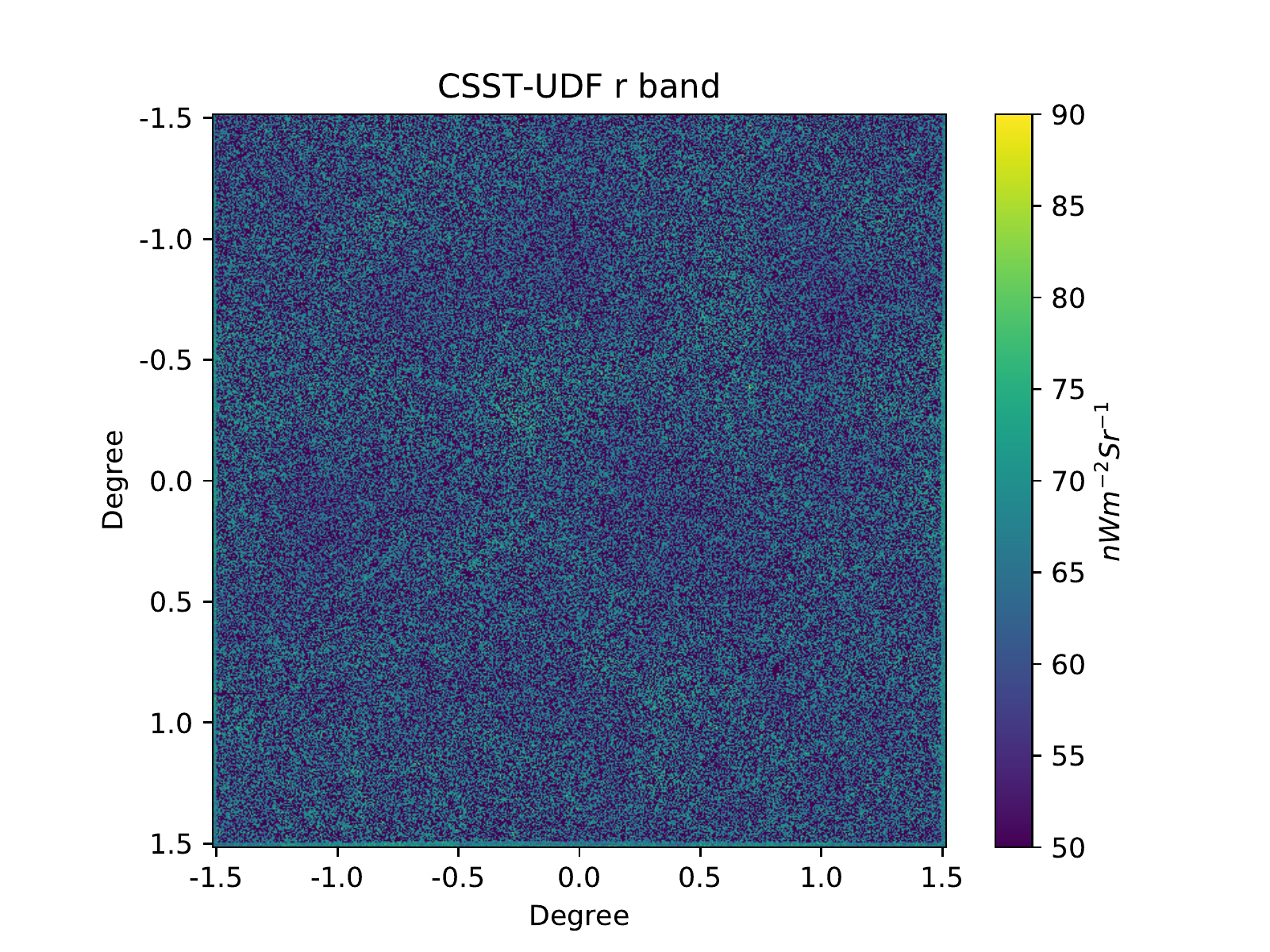}
\caption{ $\it left$: The mock background map for the $r$ band. The image shows the simulated observations in a 9 square degrees of the CSST-UDF. $\it right$: The masked image that removes the bright stars and galaxies. The mask over the whole area removes $\sim$50\% of the pixels.}
\label{fig:map0} 
\end{figure*}

The noise we add mainly consists of three components: sky background noise, dark current noise and read noise. 
The main components of the sky background come from the zodiacal light, geocoronal airglow line, reflected light from the Earth and the Moon (earthshine). Since the geocoronal airglow lines are mainly in UV bands and the Moon's contribution is small (unless the limb angle is less than $10^{\circ}$), we only consider the zodiacal light and earthshine components.
We set the sky background noise to satisfy the Poisson distribution, and estimate the sky background based on the throughput of the CSST filters and the measurements of the zodiacal light and earthshine for "LOW-SKY" background case given in \cite{Ubeda12}.
The sky background intensity as a function of wavelength is shown in Figure~\ref{fig:skybackground}, assuming the targets have large limb angles ($\ge 60^{\circ}$).
We find that the expected value of sky background are 0.084, 0.090, 0.055, and 0.016 ${\rm ADU/s/pixel}$ for the $r$, $i$, $z$ and $y$ band, respectively. The detector dark current (Poisson distribution) and read noise (Gaussian distribution) are 0.017 and 5.5 ${\rm ADU/s/pixel}$, respectively. From the above analysis, we find that the total noise of each band satisfies the Gaussian distribution.

\begin{figure*}
\centering
\includegraphics[width=1.\columnwidth]{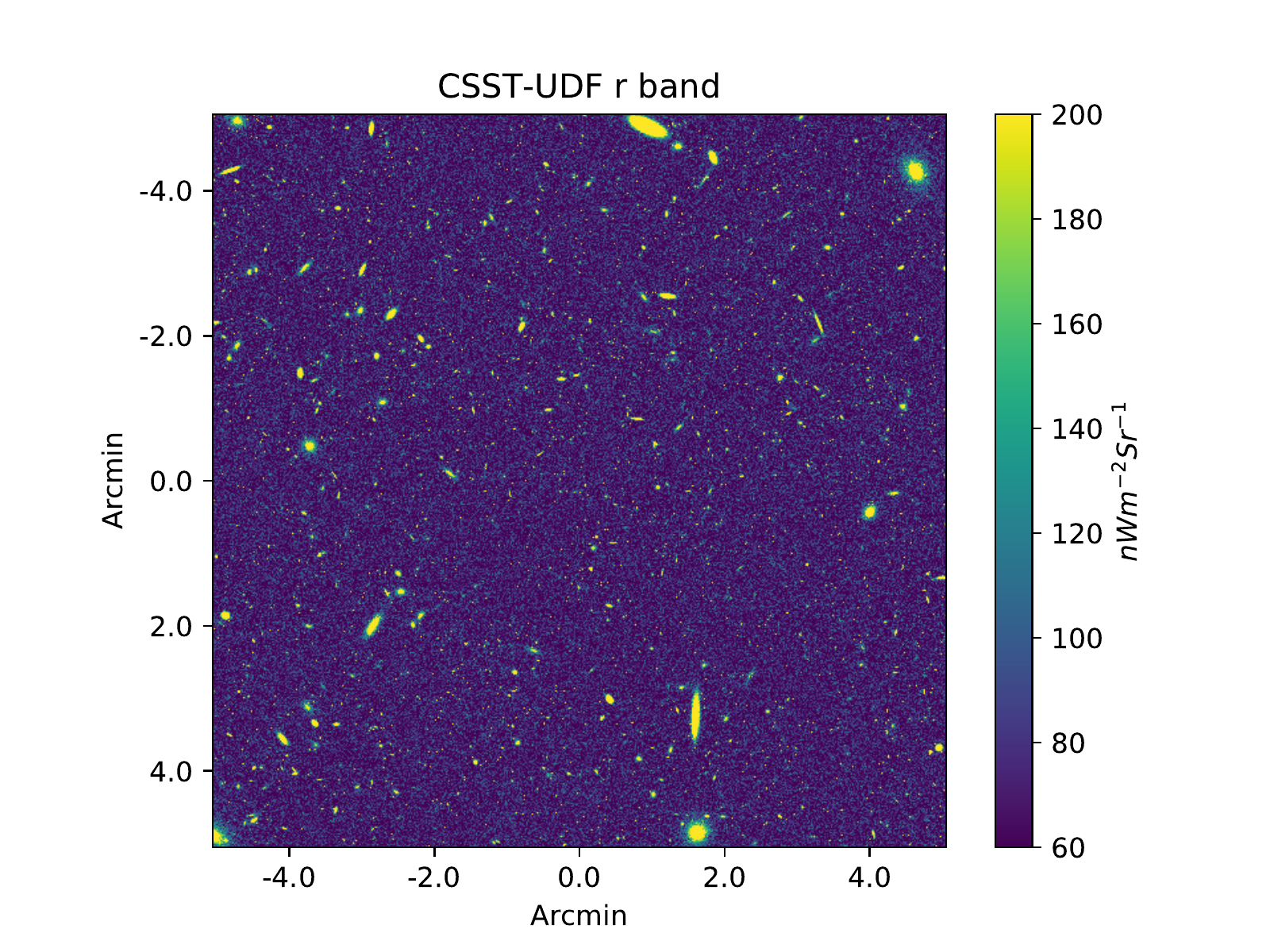}
\includegraphics[width=1.\columnwidth]{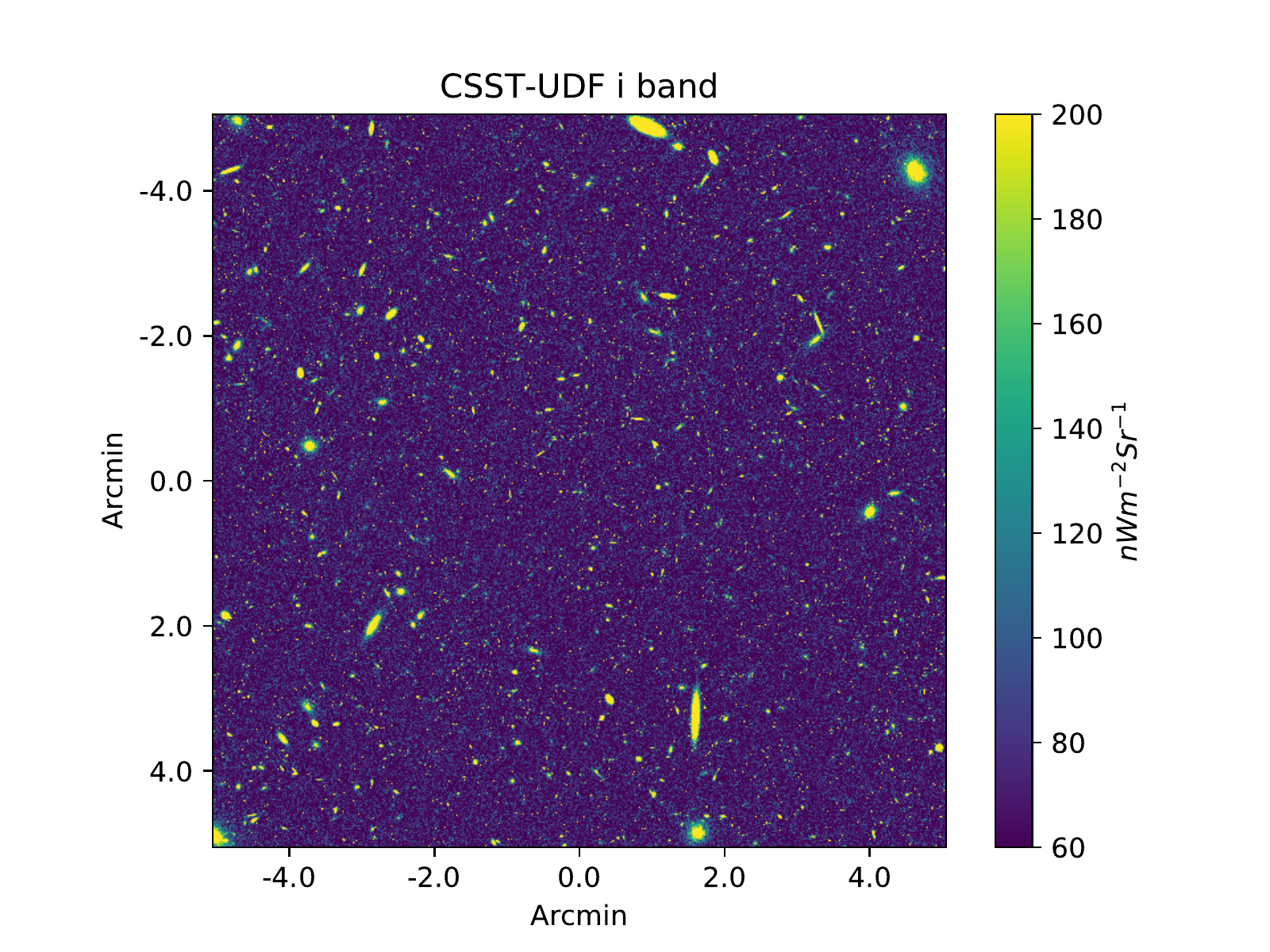}
\includegraphics[width=1.\columnwidth]{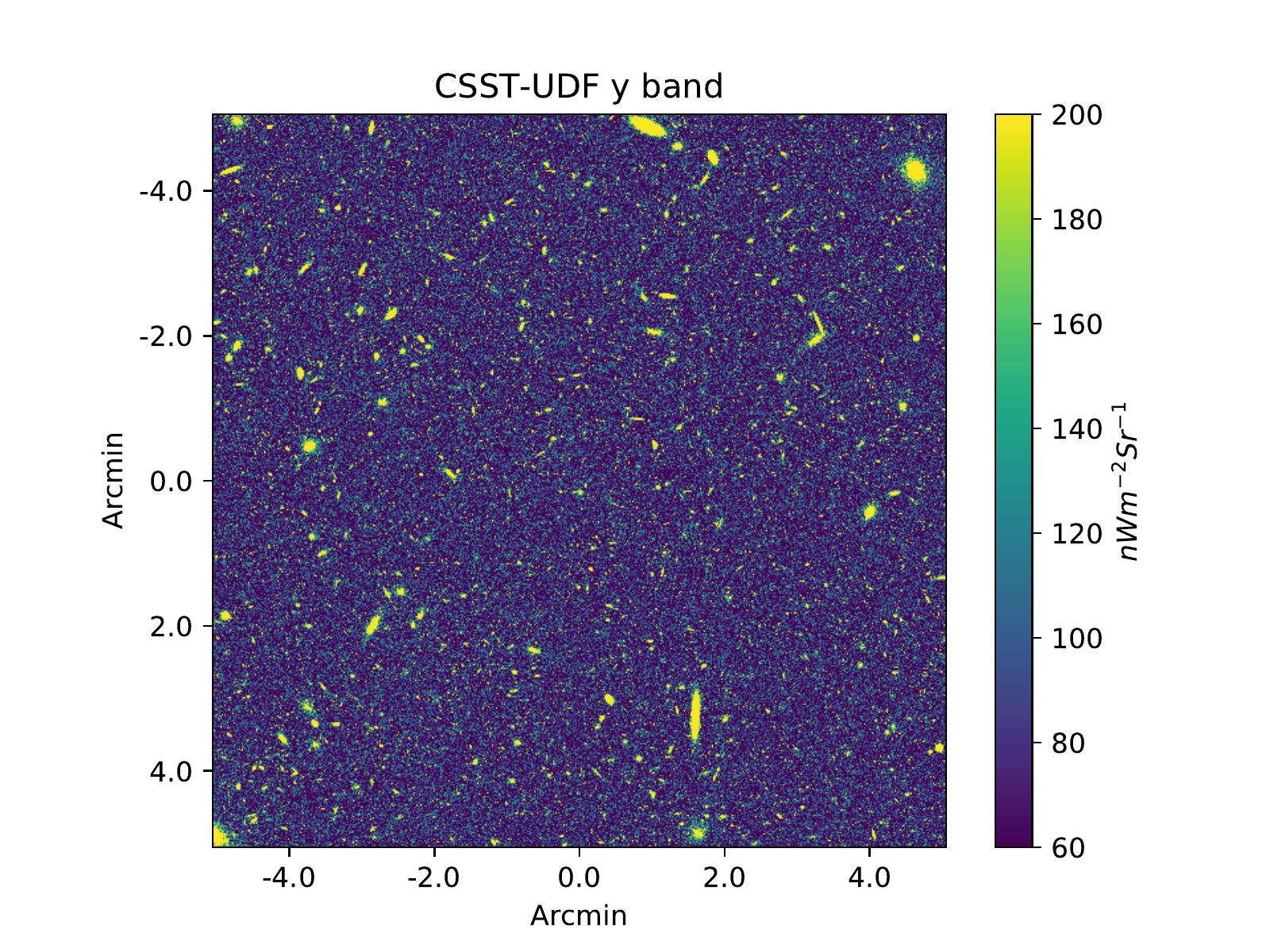}
\includegraphics[width=1.\columnwidth]{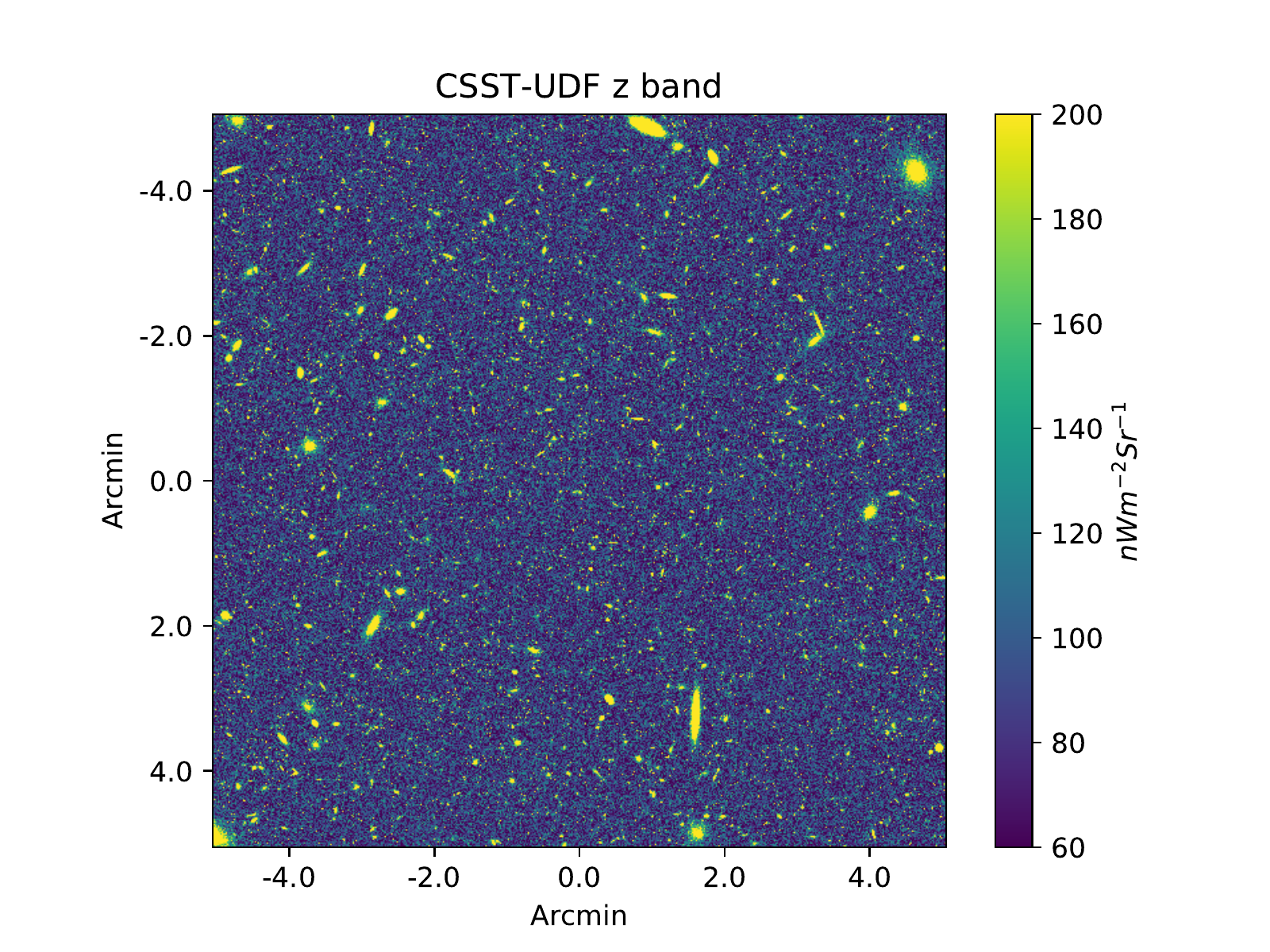}
\caption{ The zoom-in images (roughly 0.03 square degrees) of the mock CSST-UDF maps for the $r$, $i$, $z$ and $y$ bands, respectively.}
\label{fig:map1}
\end{figure*}

The sky maps are generated by merging the tiles. For the 9 ${\rm deg}^2$ CSST-UDF, we scan the field 60 times and make sure that it is fully covered each time. The scan pattern is the same in every scan, and the pointing error is $<3''$. An example of the scan pattern is shown in Figure~\ref{fig:ccd}. We present 8 tiles from two scans, here each scan contains 4 tiles, and these tiles are presented by red and green quadrilaterals. We can find that different scans in the same pointing cover the same area (the 4 tiles in the middle), and the width of the overlapping of the adjacent tiles is about $10''$. In each band, we generate tiles in the above scanning pattern based on GALSIM software, then we subtract the average value of the total noise from the tiles. Here the calibration error of these tiles is about 1\%. We generate the self-calibrated mosaics by merging all tiles for the $r$, $i$, $z$ and $y$ band, respectively. The final map at $r$ band is showed in the left-hand panel of Figure~\ref{fig:map0}. We zoom in the images of final maps to show more detailed information, and present an area of about 0.03 square degrees in Figure~\ref{fig:map1}. 

\subsection{Mask}
\label{subsec:mask}
It is necessary to remove the contamination by masking bright sources in the analysis of COB and CNIRB intensity fluctuations. The usual way is to apply a flux cut \citep{Amblard11,Thacker13,Cao20}. This method requires removing all pixels with a flux higher than the value of the flux cut, and convolve them with the PSF. Here, we use a 3$\sigma$ flux cut to remove the stars and galaxies that are presented in the mock galaxy catalog, and use the Fourier transform of the mask to calculate the weighting function, which is used for the power spectrum measurement. Then we compute the angular power spectra by using standard Fourier transform techniques on the masked maps, so that the data should be as free from foreground contamination as possible. We also mask the areas that do not contain any observed data or are at the edge of the field. Finally, we combine the masks into a final mask map, and show the masked background map for the $r$ band in the right panel of Figure~\ref{fig:map0}. We can find that the number of pixels used for calculating angular power spectrum is about 50\% of the total. %The masked maps of the previously mentioned 0.03 square degree area are shown in Figure~\ref{fig:map2}.

%\begin{figure*}
%\centering
%\includegraphics[width=1.\columnwidth]{mapr_zoom_m.pdf}
%\includegraphics[width=1.\columnwidth]{mapi_zoom_m.pdf}
%\includegraphics[width=1.\columnwidth]{mapz_zoom_m.pdf}
%\includegraphics[width=1.\columnwidth]{mapy_zoom_m.pdf}
%\caption{ 
%The zoomed in masked maps, roughly 0.03 square degrees. In this work, we mask the bright stars and galaxies and the pixels that do not contain measured data. The units of the maps are $\rm nWm^{-2}Sr^{-1}$.}
%\label{fig:map2} 
%\end{figure*}

\section{measurements of power spectra}
\label{spectrum}

The angular power spectra of the COB and CNIRB $C_{\ell}$ can be calculated by \citep{Cooray12},
\begin{equation}
 C_{\ell_{i}}=\frac{\sum_{\ell_{1}}^{\ell_{2}} w\left(\ell_{x}, \ell_{y}\right) \widetilde{\mathcal{M}}\left(\ell_{x}, \ell_{y}\right) \widetilde{\mathcal{M}}^{*}\left(\ell_{x}, \ell_{y}\right)}{\sum_{\ell_{1}}^{\ell_{2}} w\left(\ell_{x}, \ell_{y}\right)}.
\end{equation}
here $\ell_i=(\ell_1+\ell_2)/2$ is the $\ell$-mode between $\ell_1$  and $\ell_2$ in the $i$-th bin, where $\ell_1^2 < \ell_x^2 +\ell_y^2 <\ell_2^2$. $w\left(\ell_{x}, \ell_{y}\right)$ is a weighting function in Fourier space, which is obtained from the Fourier transform of mask. $\widetilde{\mathcal{M}}\left(\ell_{x}, \ell_{y}\right)$ is the two-dimensional Fourier transform of the masked background image.

The result obtained by direct measurement of masked map is a pseudo angular power spectrum, so a number of corrections are performed on the raw $\widetilde{C}_{\ell}$.  When the final angular power spectrum is estimated by,
\begin{equation}
\widetilde{C}_{\ell}=B^{2}_{\ell}T_{\ell} M_{\ell \ell^{\prime}} C_{\ell^{\prime}} + N_{\ell},
\end{equation}
where $C_{\ell^{\prime}}$ is the true angular power spectrum from full sky, $B_{\ell}$ is the beam transfer function used to correct the deviation from instrument beam in power spectrum measurement, $T_{\ell}$ is the correction of fictitious information from the map-making process, $M_{\ell \ell^{\prime}}$ is the mode coupling matrix, which can correct the errors introduced by the mask, and $N_{\ell}$ is the noise from instrumental, zodiacal light and earthshine.

\subsection{Beam Correction}

Due to the detector resolution limits, there is a non-negligible drop in the raw angular power spectrum, especially at small scales $\ell\gtrsim10^5$. We assume that all the beams of CSST are symmetric 2D-Gaussian distribution in each band. The raw power $\widetilde{C}_{\ell}$ can be corrected by a beam function $B_{\ell}$ \citep{Amblard11}. So, we can write $C_{\ell}=\widetilde{C}_{\ell} / B_{\ell}^2$, and the beam function of a symmetric 2D-Gaussian beam can be expanded as,
\begin{equation}
B_{\ell}= {\rm exp}( -\ell^2 \sigma_{\rm beam}^2/2).
\label{eq:beam}
\end{equation}
Here $\sigma_{\rm beam}$ is the standard deviation of the Gaussian beam, and is defined as $\sigma_{\rm beam} = \theta_{\rm FWHM}/\sqrt{8\ln2}$. $\theta_{\rm FWHM}$ is the full width at half maxima (FWHM) of the instrument in radian, and the FWHM values are $0.177''$, $0.190''$, $0.217''$ and $0.217''$ for the maps in the $r$, $i$, $z$ and $y$ bands, respectively, and the systematic variations are about 5\%.

\begin{figure}
\centering
\includegraphics[width=1.07\columnwidth]{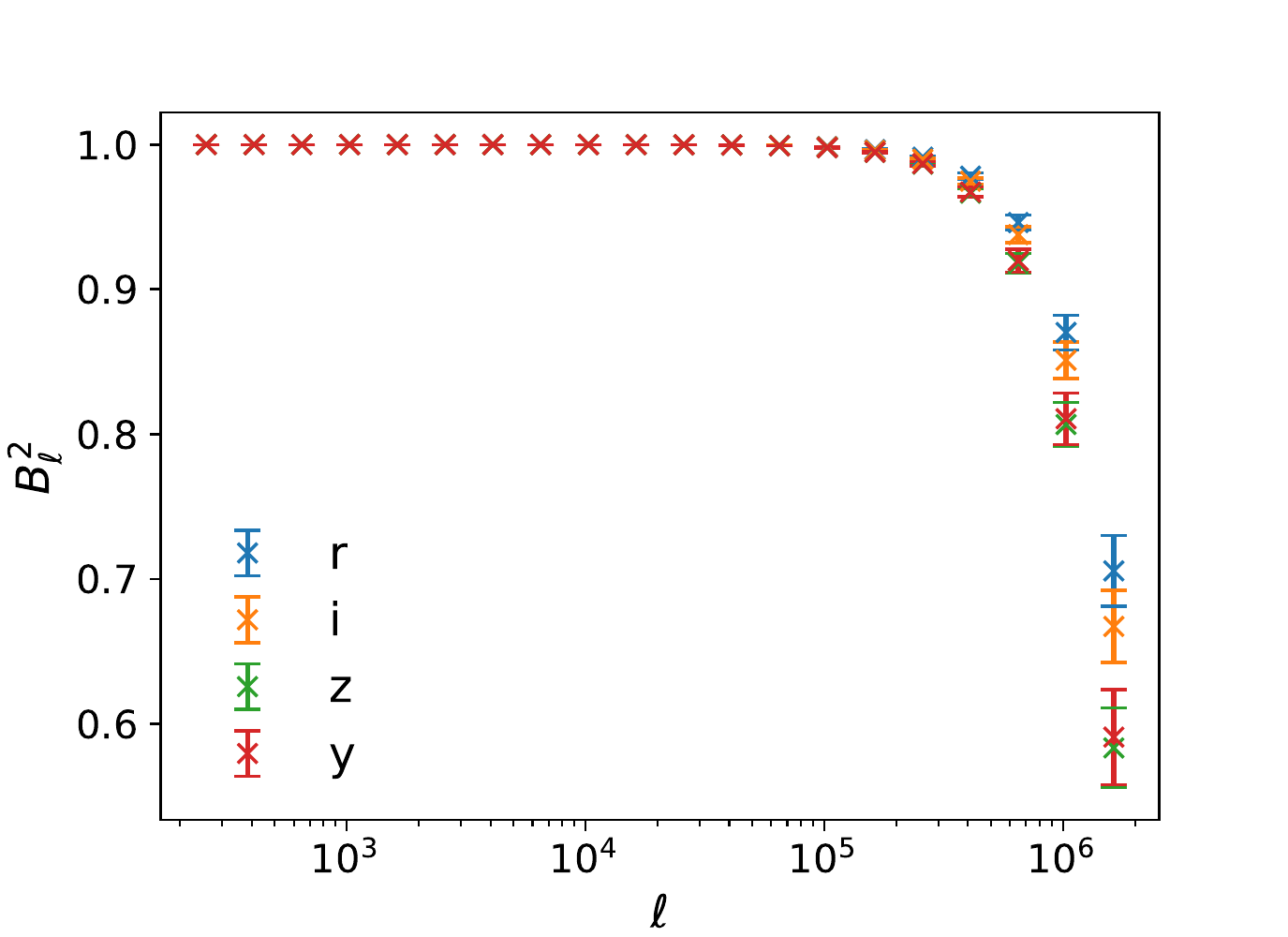} 
\caption{The beam functions in the multipole space $\ell$ for the CSST. We show the functions for the $r$, $i$, $z$ and $y$ bands with blue, orange, green and red cross, respectively. These data points and uncertainties are calculated from 100 simulations of Gaussian realizations.}
\label{fig:beam}
\end{figure}

To measure the uncertainty in the beam function we should make multiple Gaussian simulations of the PSF and calculate the power spectrum of the PSF. Then we can obtain the uncertainty $\delta B_{\ell}^2$ by,
\begin{equation}
\delta B_{\ell}^2=\frac{\delta C_{\ell}^{\rm PSF}}{C_{\ell}^{\rm point}},
\end{equation}
where, $\delta C_{\ell}^{\rm PSF}$ is the variance of the differences from the various simulations of the PSF in the $\ell$-bin, and $C_{\ell}^{\rm point}$ is the power spectrum of a point source. We show the beam transfer functions for the $r$, $i$, $z$ and $y$ bands with blue, orange, green and red cross in Figure~\ref{fig:beam}, respectively, and the uncertainties are generated from 100 simulations of Gaussian realizations.

\subsection{Mode Coupling Correction}

\begin{figure}
\centering
\includegraphics[width=1.1\columnwidth]{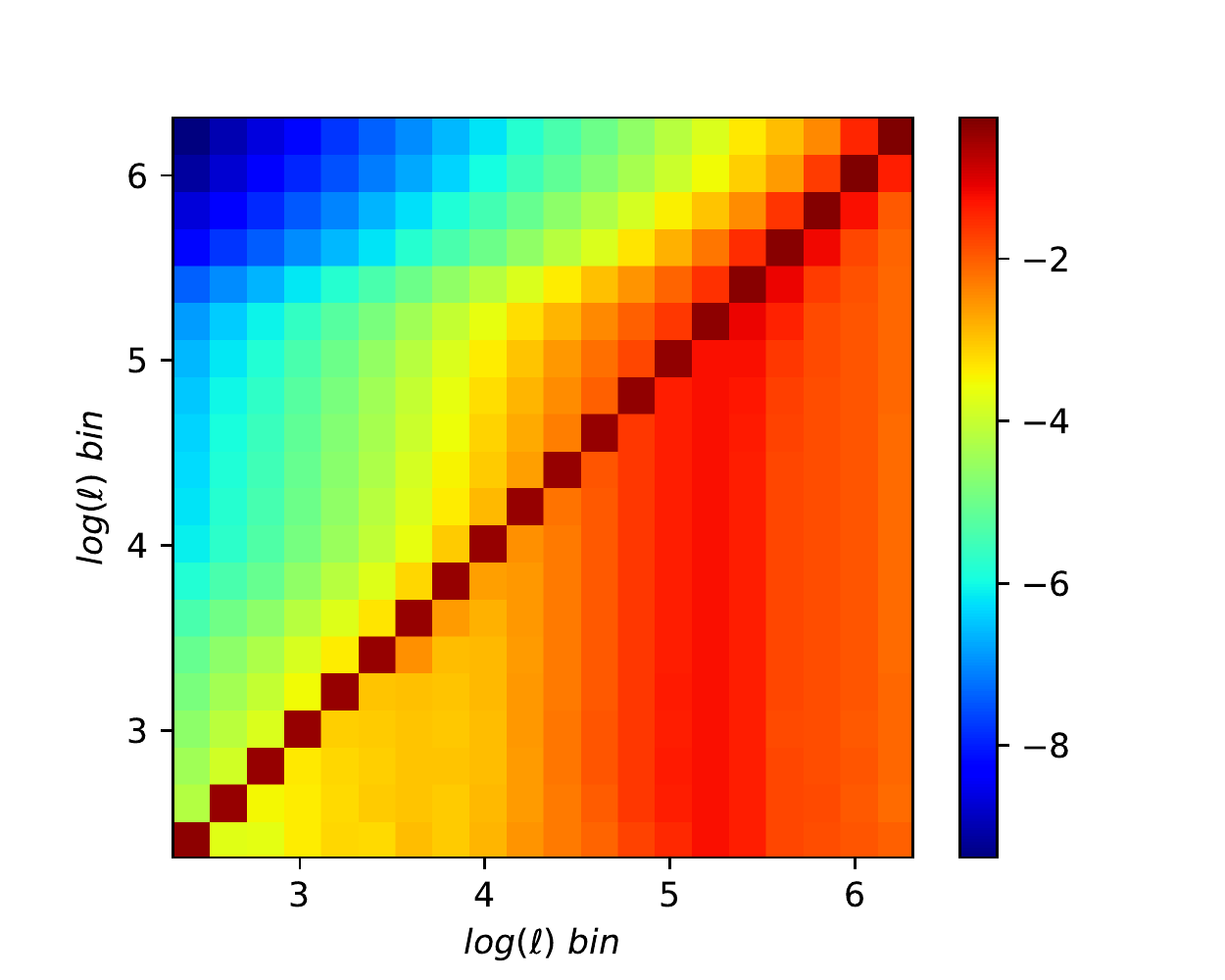}
\caption{The mode coupling matrix $M_{\ell\ell^{\prime}}$ for the $r$ band. The $\ell$ modes are equally divided into 20 bins from 200 to 2,000,000 in logarithmic scale, and the coordinates of the color bar is also in logarithmic scale. We can find that the effect between adjacent bins is greater, and high $\ell$ modes (small scale) has a greater effect than low $\ell$ modes (large scale). }
\label{fig:mll}
\end{figure}

When removing the contamination from the brighter sources in our power spectrum measurement, we mask some pixels from the sky maps. This will break the high $\ell$ modes into low $\ell$ modes, which leads to a lot of fictitious information converting into the power spectrum when performing Fourier transform. As a result, the profile of the measured angular power spectrum will change after using the mask maps. Here, we use a mode coupling matrix $M_{\ell \ell^{\prime}}$ to correct this change \citep{Cooray12,Thacker15}. The mode coupling matrix is generated obtained based on the $\rm MASTER$ model \citep{Hivon02}, and we calculate them by analyzing the effects of the mask at $\ell$-mode. 
%For a masked sky map, the process of calculating the mode coupling matrix $M_{\ell \ell^{\prime}}$ consists of four main steps. First, we need generate a simulation realization from a pure tone power spectrum (where $C_{\ell^{\prime}} = 1$ if $\ell^\prime=\ell$, otherwise $C_{\ell^{\prime}} = 0$), and mask the realization using the real space mask map. Second, we calculate the raw power spectra of these masked maps with a pure tone power spectrum, and the results can show how the mask mixes the power from $\ell-$mode into other modes. Third, we repeat the above two steps and run the simulation 100 times for $\ell-$mode, then average the power spectrum as a row of the mode coupling matrix $M_{\ell \ell^{\prime}}$, so we can express as $M_{\ell}=\left\langle{\widetilde{C}}_{\ell^{\prime}}\right\rangle$. Finally, we repeat the above process for all $\ell$ bins. 
For a masked sky map, the mode coupling matrix $M_{\ell \ell^{\prime}}$ is calculated by a simulation procedure. Then we calculate the inverse of the mode coupling matrix to correct for the effect of the mask. The real power spectrum is  $C_{\ell^{\prime}}= M_{\ell \ell^{\prime}}^{-1}\widetilde{C}_{\ell}$. We show the mode-coupling matrix at $i$ band in Figure~\ref{fig:mll}. 
 
Masking reduces the coverage of the unmasked sky and results in an increase in cosmic variance, but the mode-coupling matrix does not restore the variation from cosmic variance. We combine the errors from masking and correcting into the cosmic variance, and we obtain the cosmic variance $\delta C_{\ell}$ of the auto-power spectrum by \citep{Cooray12,Zemcov14,Mitchell-Wynne15,Thacker15}
\begin{equation}
\delta C_{\ell}=\sqrt{\frac{2}{f_{\rm sky}(2 \ell+1) \Delta \ell}}C_{\ell},
\end{equation}
where, $f_{\rm sky}$ is the fraction of the unmasked areas of all sky, $ \Delta \ell$ is the width of the $\ell$-bin.

\subsection{Transfer Function}
\label{subsec:trans_func}

Because of the limitation of single CCD observation range, we have to obtain the large sky filed image by combining the tiles, then combine those tiles into the final sky map. Since the calibrated tiles still have errors compared to the real sky, and the values of the errors between the different tiles are different, this fictitious information will be converted to the final power spectrum by performing a Fourier transform. Through the map-making procedure, the shape of power spectrum inevitably changes, so the effect of the map-making procedure must be quantified and corrected. 

\begin{figure}
\centering
\includegraphics[width=1.07\columnwidth]{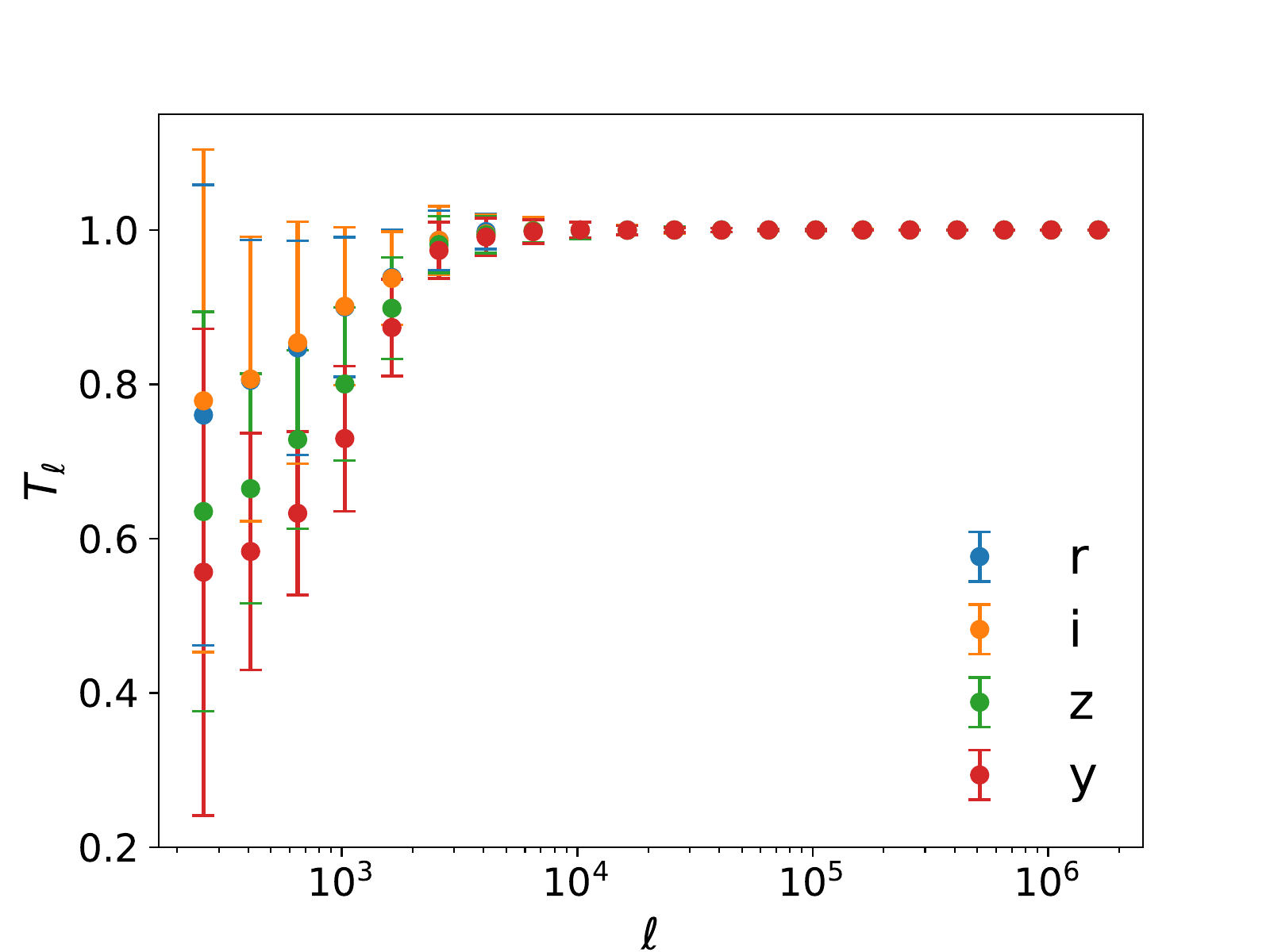} 
\caption{The map-making transfer functions at $r$ (0.620$\rm \mu m$, blue points), $i$ (0.760$\rm \mu m$, orange points), $z$ (0.915$\rm \mu m$, green points), and $y$ band (0.965$\rm \mu m$, red points), respectively. These data points and uncertainties are calculated from 100 simulations of Gaussian realizations. Due to a similar scanning pattern at the four bands, the values of deviation are almost same at large scales (low $\ell$ modes), and the transfer functions are approximately equal to 1 at small scale.}
\label{fig:tran}
\end{figure}

Here, we rebuild the map to study the effects of the map-making procedure on the angular power spectrum. First, we use a given power spectrum $C_{\ell}^{\rm in}$ to generate randomly a Gaussian realization that is similar to our simulated observation in size, pixel scale and astrometry. Next, we break the simulated sky map into the tiles using the same CSST scan mode, and add a calibration error to each tile according to the instrument parameters. Finally, we merge the new tiles into final sky map, and we calculate the raw power spectra $C_{\ell}^{\rm out}$ of the new maps, and defined the map-making transfer function as $T_{\ell} =C_{\ell}^{\rm out}/C_{\ell}^{\rm in}$. The transfer function is independent of the shape and amplitude of the input power spectrum \citep{Cooray12}. In order to avoid the potential error caused by the use of a specific input power spectrum, we repeat all the steps above using different input power spectrum, and take the average as the final map-making transfer function, and calculate the standard deviation of the 100 simulations as the uncertainties. The map-making transfer functions for the $r$, $i$, $z$ and $y$ bands are show in Figure~\ref{fig:tran}. We can see that the $T_{\ell}$ is almost the same for the four bands, and the attenuation feature is mainly due to their scanning patterns are very similar \citep{Viero13,Thacker15}. The transfer functions are approximately equal to 1, and the uncertainties are very small at small scales.

\subsection{Noise}
\label{subsec:noise}

\begin{figure}
\centering
\includegraphics[width=1.0\columnwidth]{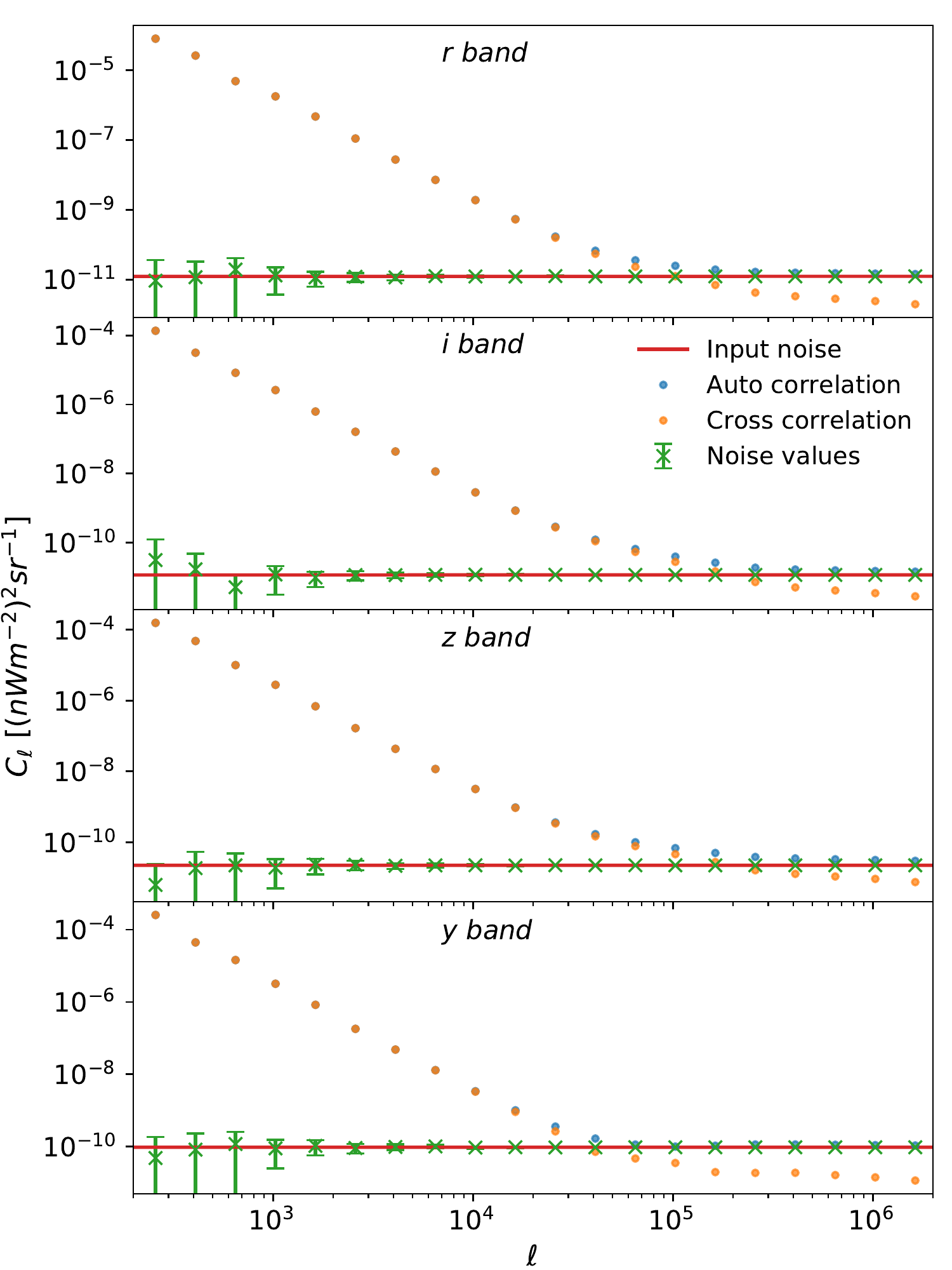} 
\caption{The noise power spectrum for each band. The blue data points show the auto-power spectra, which is a combination of the sky signal and the instrumental noise. We estimate the raw power spectra (orange points) by dividing all the tiles equally into two epochs. The green cross with error bars show the instrumental noise power spectra, which are obtained by calculating the difference between of the auto- and cross-power spectra. The solid red lines represent the curves of input noise added to the simulated background.}
\label{fig:noise}
\end{figure}

The instrumental noise has two components: detector dark current and read noise. The value of dark current at each pixel satisfies the Poisson distribution, and its expected value is 0.017 ${\rm ADU/s/pixel}$. The read noise satisfies a Gaussian distribution with a root mean square of 5.5 ${\rm ADU/s/pixel}$. In this work, we can regard the sum of instrumental noise, zodiacal light and earthshine as a white noise, and minimize the contamination from these noise by cross-correlating two different maps in a field. Since each area of the CSST-UDF has been observed 60 times, we can divide all the tiles equally into two epochs with 30 observations each. Then we obtain the power spectrum of noise by calculating the difference between auto-correlation and cross-correlation of the two epochs. The noise power spectra are presented in Figure~\ref{fig:noise}. We can find that the noise follows a white-noise power spectrum in each band.

\section{Results and analysis of power spectrum}
\label{spec_result}

In this section, we present the measured final power spectra and compare them with the fiducial value. Then we fit the measured power spectra using the theoretical model and perform a MCMC analysis to constrain the model parameters.

\subsection{Final Power Spectra}
\label{subsec:final_power}

\begin{figure*}
\centering
\includegraphics[width=1.0\columnwidth]{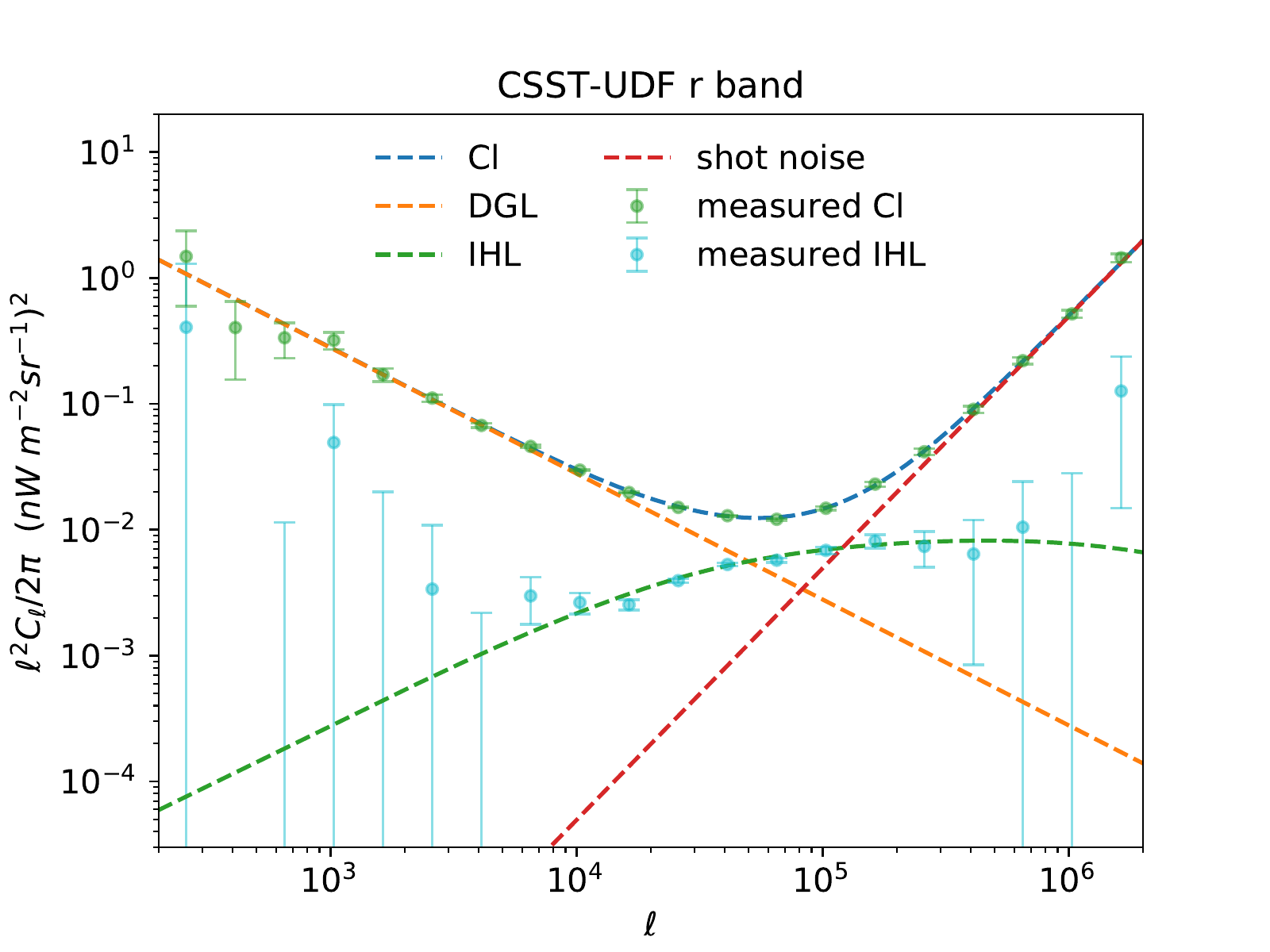}
\includegraphics[width=1.0\columnwidth]{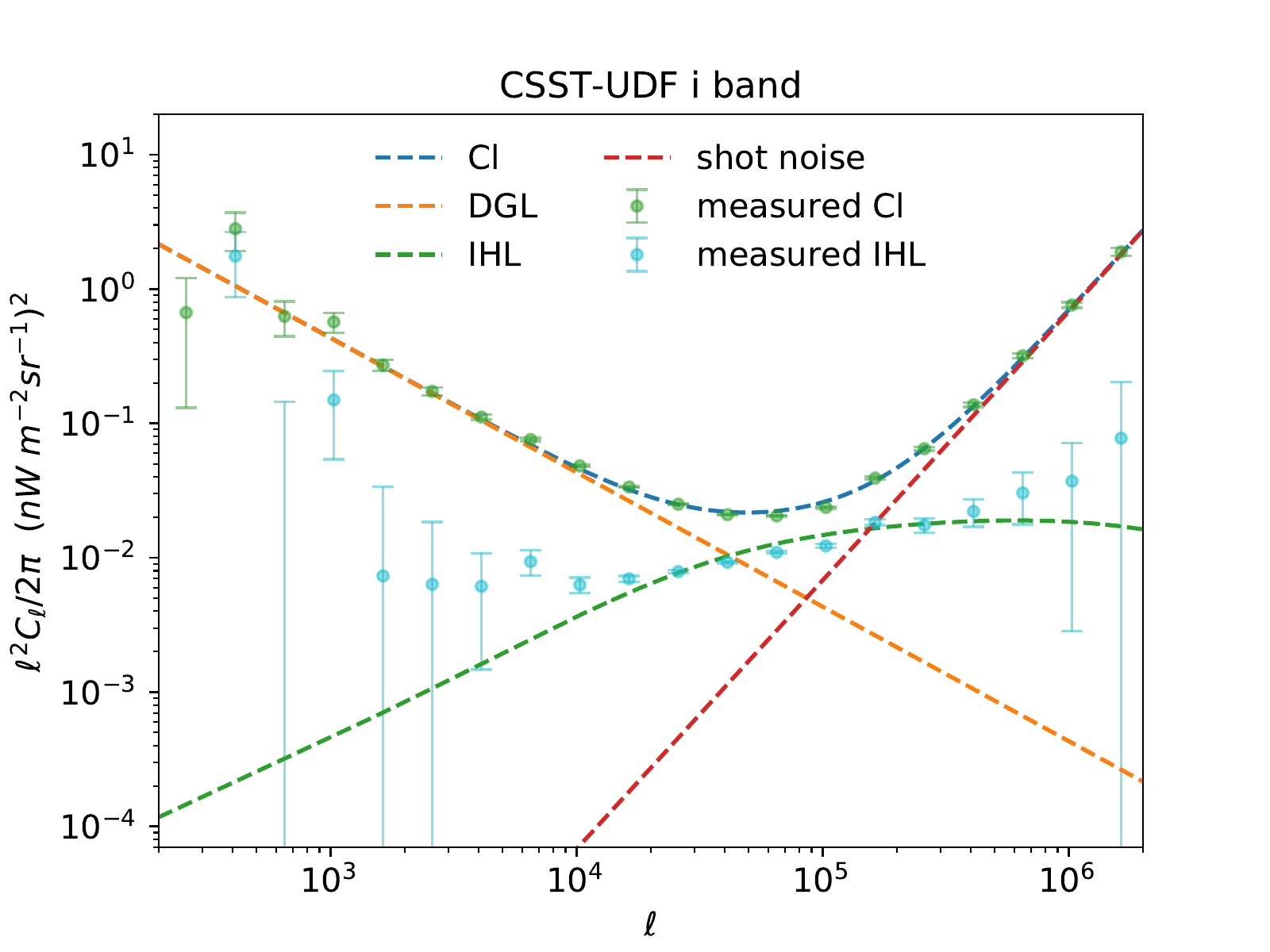}
\includegraphics[width=1.0\columnwidth]{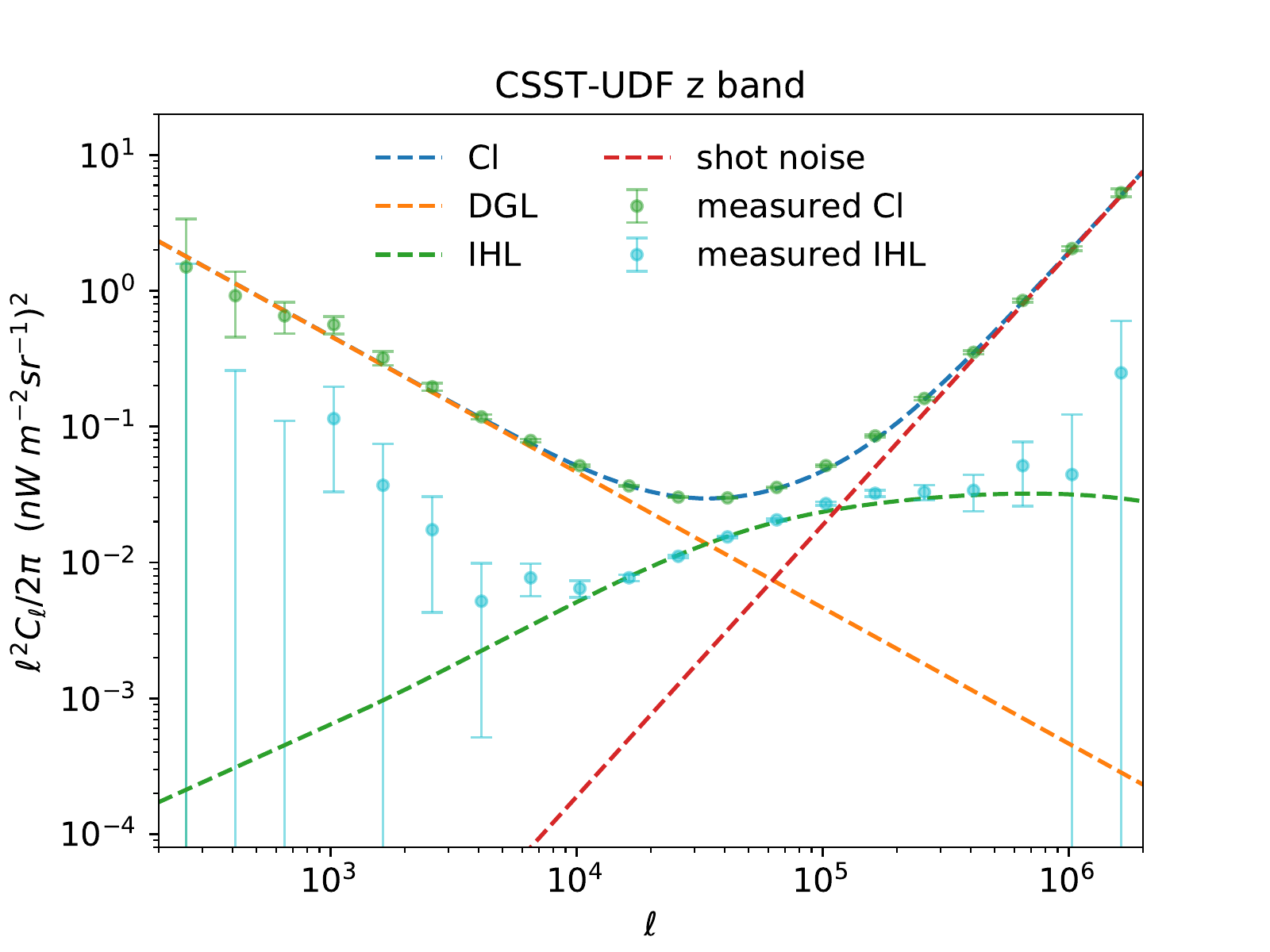}
\includegraphics[width=1.0\columnwidth]{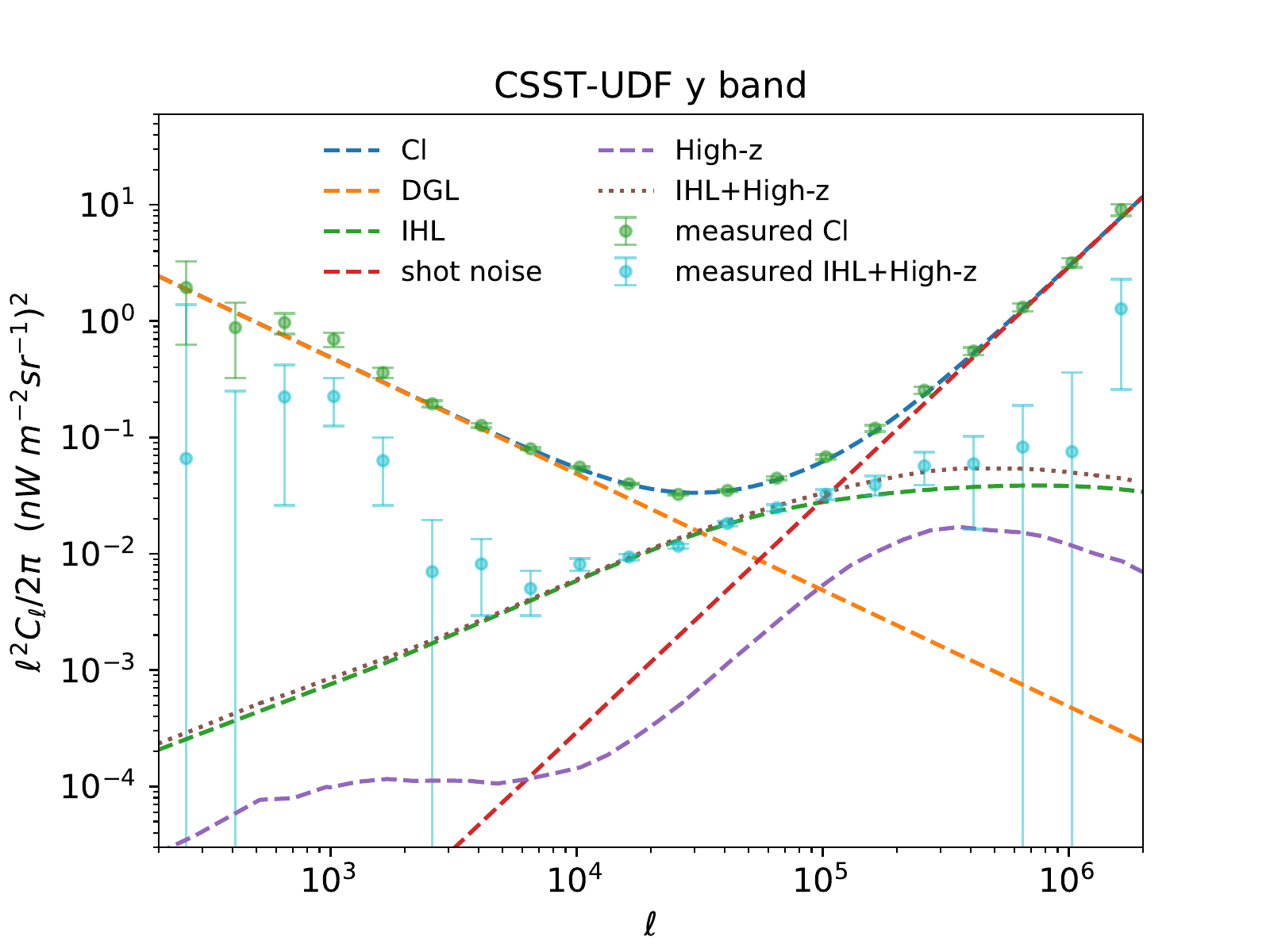}
\caption{ Multi-wavelength auto angular power spectra of CSST surveys at at $r$, $i$, $z$ and $y$ band. We show the measured total power spectra in green data points, and the ones subtracting the DGL and shot noise in cyan data points. The error bars are 1$\sigma$ uncertainties which are obtained by calculating the square root of errors from the beam function, cosmic variance, map-making, and calibration. The blue, orange, green and red dashed lines are the fiducial total theoretical power spectrum and its three components: DGL, IHL and shot noise. Note that the signal from $z>6$ high-redshift galaxies (purple dashed lines) is contained only in the $y$ band. The DGL, IHL and shot noise components play a dominant role in the total power spectrum at low multipoles (the scales greater than a few arcminutes), intermediate multipoles (the scales of about one arcminute) and high multipoles (the scales less than one arcminute), respectively.}
\label{fig:llcl}
\end{figure*}

%We compute the power spectra of the COB and CNIRB using the simulated data from the CSST-UDF. We generate the mask map, and remove the brighter stars and galaxies. Then we analyze the differences between auto-correlation and cross-correlation between the two epochs to remove the noise. We use the standard Fourier Transform techniques to compute the power spectrum. After the raw spectra are obtained from the mock sky maps, they are corrected for beam function, mode coupling matrix and transfer function. We then investigate if the current CSST instrument parameters definition can provide accurate COB and CNIRB intensity fluctuations results that can achieve the science requirement in certain parameters ranges. 

In Figure~\ref{fig:llcl}, we show the final multi-wavelength auto angular power spectra of the CSST-UDF surveys as $\ell^2C_{\ell}/2\pi$ in the $r$, $i$, $z$ and $y$ bands, respectively. Note that the signal from $z>6$ high-$z$ galaxies (purple dashed lines) is contained only in the $y$ band. Here, the corrected total power spectra of the measured background intensity fluctuations are presented as green data points over a larger scale range from $\ell=$200 to 2,000,000. The power spectra after removing DGL and shot noise are presented as cyan data points, and the plotted error bars are 1$\sigma$ uncertainties, which are obtained by calculating the square root of errors from the beam function, cosmic variance, map-making, and calibration. We use the blue, orange, green and red dashed lines to show the fiducial total theoretical power spectrum and its three components, i.e. the DGL, IHL and shot noise, respectively. 

As can be seen, the DGL component is the dominant contributor at low $\ell$-mode (on scales greater than a few arcmin). Due to the large scale of the CSST-UDF, we expect to obtain more accurate measurements of DGL composition. The shot noise plays a dominant role in the total power spectrum at high $\ell$-mode (on scales less than one arcmin), and the IHL becomes the dominant component at intermediate $\ell$-mode. In this work, we can find that the date points of the measured IHL power spectrum is very similar to the input values (green dashed lines) at 10,000$\ \le\ell \le\ $400,000 that is useful for astrophysical studies, and their signal-to-noise ratios (SNRs) is larger than 8, which is accurate enough for analysis. Due to the influence of the map-making procedure and the limitation of the unmasked observation area, our results have a large deviation at the large scale, but they are basically within 1-$\sigma$ uncertainty. According to the analysis of the section \S\ref{spectrum}, we know that the deviation at the small scale is mainly due to the influence of beam, shot noise, and instrumental noise. In Figure~\ref{fig:llcl}, we can find that the shot noise amplitude is a constant and dominant at small scale, we obtain the values and uncertainties of shot noise through the fitting process.

%From what has been discussed above, we find that the current CSST can provide good observation and analysis results in certain instrument parameters and scanning pattern, that can meet the science requirement.

\subsection{Fitting results}
\label{subsec:fit_result}

\begin{figure*}
\centering
\includegraphics[width=2.\columnwidth]{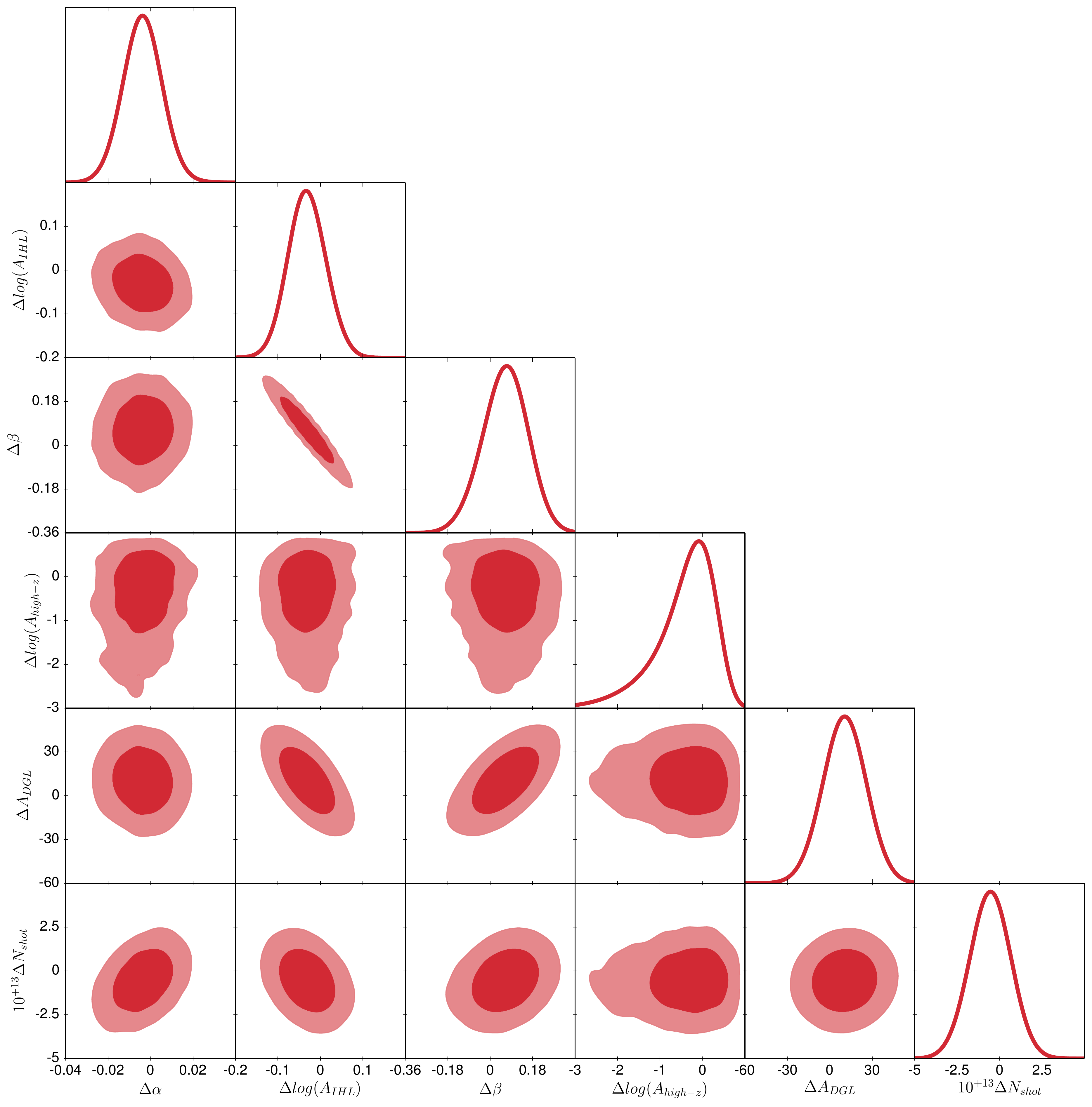}
\caption{ The contour maps with 1$\sigma$ and 2$\sigma$ C.L. of the free parameters subtracting the fiducial values, i.e. $\Delta \alpha$, $\Delta \beta$, $\Delta \log_{10}(A_{\rm IHL})$, $\Delta \log_{10}(A_{\rm high-z})$ in the $y$ band, and $\Delta A_{\rm DGL}$ and $\Delta N_{\rm shot}$ in the $r$ band. The solid lines are the 1-D PDFs of the parameters. Note that we only show the results of $\Delta A_{\rm DGL}$ and $\Delta N_{\rm shot}$ in the $r$ band, since they are quite similar for the other bands.}
\label{fig:fit} 
\end{figure*}

\begin{table*}
\begin{center}
\caption{\label{tab:result} The best-fits and errors of the parameters in the model from the MCMC fitting.}
\small
\begin{tabular}{c c c c}
\hline\hline
\rule[-2mm]{0mm}{6mm}
Parameter & $\rm input$ & $\rm Best fit$ & $\rm Prior\ min,max$\\
\hline
\rule[-2mm]{0mm}{6mm}
$\log_{10}(A_{\rm IHL})$      &$-3.23$ &$-3.26^{+0.04}_{-0.04}$&$-6, 10$\\
\rule[-2mm]{0mm}{6mm}
$\log_{10}(A_{\rm high-z})$  &$1.19$ &$1.09^{+0.40}_{-0.63}$&$-5, 7$\\
\rule[-2mm]{0mm}{6mm}
$\alpha$                               &$0.094$&$0.090^{+0.010}_{-0.009}$&$\ \ 0, 1$\\
\rule[-2mm]{0mm}{6mm}
$\beta$                                 &$1.0$&$1.06^{+0.10}_{-0.09}$&$-5, 5$\\
\rule[-2mm]{0mm}{6mm}
$A_{\rm DGL}^{r}$                &$1750.0$&$1760.4^{+15.5}_{-15.3}$&$100, 10000$\\
\rule[-2mm]{0mm}{6mm}
$A_{\rm DGL}^{i}$                 &$2710.0$&$2733.6^{+27.7}_{-27.6}$&$100, 10000$\\
\rule[-2mm]{0mm}{6mm}
$A_{\rm DGL}^{z}$                &$2910.0$&$2930.1^{+32.4}_{-32.1}$&$100, 10000$\\
\rule[-2mm]{0mm}{6mm}
$A_{\rm DGL}^{y}$                &$3050.0$&$3074.5^{+37.1}_{-37.2}$&$100, 10000$\\
\rule[-2mm]{0mm}{6mm}
$N_{\rm shot}^{r}$                 &$3.12 \times 10^{-12}$&$3.07^{+0.12}_{-0.12} \times 10^{-12}$&$1\times 10^{-12},\ 4\times 10^{-11}$\\
\rule[-2mm]{0mm}{6mm}
$N_{\rm shot}^{i}$                 &$4.29  \times 10^{-12}$&$4.20^{+0.14}_{-0.14} \times 10^{-12}$&$1\times 10^{-12},\ 4\times 10^{-11}$\\
\rule[-2mm]{0mm}{6mm}
$N_{\rm shot}^{z}$                &$1.19 \times 10^{-11}$&$1.21^{+0.03}_{-0.03} \times 10^{-11}$&$1\times 10^{-12},\ 4\times 10^{-11}$\\
\rule[-2mm]{0mm}{6mm}
$N_{\rm shot}^{y}$                 &$1.84 \times 10^{-11}$&$1.76^{+0.09}_{-0.09} \times 10^{-11}$&$1\times 10^{-12},\ 4\times 10^{-11}$\\
\hline
\end{tabular}
%\tablecomments{Superscript and subscript represent 68$\%$ confidence level.}
\end{center}
\end{table*}

After we obtain four angular power spectra from  the CSST-UDF in the four bands, we perform joint fits for these data with the theoretical model described in \S\ref{model} using the MCMC method \citep{Metropolis53}. The Monte Python software \citep{Audren13} is used to calculate the acceptance probability of the MCMC chain point, and analyze the model parameters of the theoretical model. We use the $\chi ^2$ distribution to estimate the likelihood function, that we have $\mathcal{L} \sim \exp(-\chi^2_{\rm tot}/2)$. Here, the total $\chi^2_{\rm tot}=\sum \chi^2$ is obtained by the sum of the data from these four bands, and the $\chi^2$ value is calculated by
\begin{equation} \label{eq:chi2}
\chi^2=\sum_{i=1}^{N} \left(\frac{C_{\ell_i}^{\rm obs}-C_{\ell_i}^{\rm th}}{\sigma_{\ell_i}^{\rm obs}}\right)^2,
\end{equation}
where $N$ is the number of data points of power spectrum in each band, $C_{\ell_i}^{\rm obs}$ and $C_{\ell_i}^{\rm th}$ are the observed and theoretical power spectra at the $i$-th bin, respectively, and $\sigma_{\ell}^{\rm obs}$ is the error corresponding to the observed power spectrum data. 

The theoretical model contains several free parameters, including $\alpha$, $\beta$, $A_{\rm IHL}$, $A_{\rm high-z}$ (for the $y$ band), four $A_{\rm DGL}$ and four $N_{\rm shot}$, which totally has 12 free parameters. In this work, we set the flat priors of the model parameters: $\alpha \in ( 0, 1)$, $\beta \in ( -5, 5)$, $\log_{10}(A_{\rm IHL}) \in ( -6,10)$, $\log_{10}(A_{\rm high-z}) \in ( -5,7)$, $A_{\rm DGL} \in$ (100, 10,000) and $N_{\rm shot} \in ( 1\times 10^{-12}, 4\times 10^{-11})$. For each case, we run 50 chains, and each chain contains 500,000 steps. We have shown the fitting results and 68$\%$ confidence levels (C.L.) in Table~\ref{tab:result}.

The power spectrum in each band consists of 20 data points ranging $\ell$ mode from 200 to 2,000,000,  so the degree of freedom is $N_{\rm dof} = 20\times4-12= 68$. We find that the reduced chi-square $\chi^2_{\rm red}=\chi^2_{\rm min}/N_{\rm dof}$ is less than 0.5. Therefore, we can fit the data quite well, and the best-fit curves are consistent with most of the data points in 1-$\sigma$. From the above discussion, we know that the DGL and shot noise play a dominant role at large scale and small scale, respectively. The large sky area is very helpful for the accurate derivation of the DGL components, and  accurate measurement of instrument beam helps to analyze the shot noise. From Table~\ref{tab:result} we can find that, all the best-fit values of free parameters are very close to the input values within 1-$\sigma$.

We have shown the contour maps of $\Delta \alpha$, $\Delta \beta$, $\Delta \log_{10}(A_{\rm IHL})$, $\Delta \log_{10}(A_{\rm high-z})$, $\Delta A_{\rm DGL}^r$ and $\Delta N_{\rm shot}^r$ with 1-$\sigma$ and 2-$\sigma$ C.L. in the Figure~\ref{fig:fit}. We find that, the fitting results of $\alpha$, $\beta$, $A_{\rm IHL}$ and $A_{\rm high-z}$ are consistent with the results measured by \cite{Cooray12} and \cite{Mitchell-Wynne15}, and we have much smaller uncertainties since the CSST has better observation capabilities. In details, \cite{Cooray12} and \cite{Mitchell-Wynne15} did similar analysis using the data from the Spitzer Deep Wide Field Survey (SDWFS) \citep{Eisenhardt04,Ashby09} and Cosmic Assembly Near-Infrared Deep Extragalactic Legacy Survey (CANDELS) \citep{Grogin11,Koekemoer11}, respectively. The 0.606$\rm{\mu m}$, 0.775$\rm{\mu m}$ and 0.850$\rm{\mu m}$ bands they used are similar to the $r$ (0.620$\rm \mu m$), $i$ (0.760$\rm{\mu m}$) and $z$ (0.915$\rm{\mu m}$) bands in the CSST-UDF survey, respectively. Our constraints on $\alpha$ and $A_{\rm high-z}$ are similar as theirs, and the constraints on $A_{\rm IHL}$ and $\beta$ are better than theirs by factors of $\sim$3 and $\sim$4,  respectively.

\section{Summary and conclusion}
\label{conclusions}

In this work, we generate the multi-band sky maps for the CSST-UDF, which has about 9 square degree and the limiting magnitude is 28.3, 28.2, 27.6, 26.7 for point sources with 5$\sigma$ detection at $r$, $i$, $z$ and $y$ band, respectively. Then we analyze the auto power spectra of the anisotropies of COB and CNIRB to explore the observation accuracy of the CSST, and explore the implications for cosmological studies by fitting the power spectrum to the theoretical model.

We adopt a multi-component model, which involves intrahalo light, diffuse galactic light, low- and high-redshift galaxies and the instrumental and shot noise in our analysis. We generate tiles for each exposure observation in CSST scan mode based on GALSIM image simulation software. Then we merge the sky maps using the self-calibration technique, and set the CSST pixel value of $0.074''$. The raw power spectra are obtained from the masked sky maps, which have removed the pixels larger than flux cut. Beam function, transfer function and mode coupling matrix are calculated to correct the effects of instrument beam, map-making progress and mask, respectively. Then we obtain and compare the measured angular power spectra with the inputs.

We also discuss the results from the data analysis and use the multi-component model to fit the corrected power spectrum data using the MCMC method. There are 12 free parameters, and we can found that the fitting results are very well, with the reduced chi-squares is less than 0.5. Comparing to previous works, our constraints on $\alpha$ is similar as theirs, and the constraints on $A_{\rm IHL}$ and $\beta$ are better than theirs by factors of $\sim$3 and $\sim$4, respectively. This improvement indicate the CSST is powerful tool to explore the COB and CNIRB.

\section*{Acknowledgements}
%The Acknowledgements section is not numbered. Here you can thank helpful
%colleagues, acknowledge funding agencies, telescopes and facilities used etc.
Y.C. and Y.G. thank Linhua Jiang for helpful discussion. Y.C. and Y.G. acknowledge the support of NSFC-11822305, NSFC-11773031, NSFC-11633004, MOST-2018YFE0120800, 2020SKA0110402, and CAS Interdisciplinary Innovation Team. X.L.C. acknowledges the support of the National Natural Science Foundation of China through grant No. 11473044, 11973047, and the Chinese Academy of Science grants QYZDJ-SSW-SLH017, XDB 23040100, XDA15020200. This work is also supported by the science research grants from the China Manned Space Project with NO.CMS-CSST-2021-B01 and CMS- CSST-2021-A01.

%%%%%%%%%%%%%%%%%%%%%%%%%%%%%%%%%%%%%%%%%%%%%%%%%%
\section*{Data Availability}

The data that support the findings of this study are available from the corresponding author, upon reasonable request.
 
%The inclusion of a Data Availability Statement is a requirement for articles published in MNRAS. Data Availability Statements provide a standardised format for readers to understand the availability of data underlying the research results described in the article. The statement may refer to original data generated in the course of the study or to third-party data analysed in the article. The statement should describe and provide means of access, where possible, by linking to the data or providing the required accession numbers for the relevant databases or DOIs.



\bibliographystyle{mnras}
\bibliography{CSST_background.bib} % if your bibtex file is called example.bib

\begin{thebibliography}{}
\makeatletter
\relax
\def\mn@urlcharsother{\let\do\@makeother \do\$\do\&\do\#\do\^\do\_\do\%\do\~}
\def\mn@doi{\begingroup\mn@urlcharsother \@ifnextchar [ {\mn@doi@}
  {\mn@doi@[]}}
\def\mn@doi@[#1]#2{\def\@tempa{#1}\ifx\@tempa\@empty \href
  {http://dx.doi.org/#2} {doi:#2}\else \href {http://dx.doi.org/#2} {#1}\fi
  \endgroup}
\def\mn@eprint#1#2{\mn@eprint@#1:#2::\@nil}
\def\mn@eprint@arXiv#1{\href {http://arxiv.org/abs/#1} {{\tt arXiv:#1}}}
\def\mn@eprint@dblp#1{\href {http://dblp.uni-trier.de/rec/bibtex/#1.xml}
  {dblp:#1}}
\def\mn@eprint@#1:#2:#3:#4\@nil{\def\@tempa {#1}\def\@tempb {#2}\def\@tempc
  {#3}\ifx \@tempc \@empty \let \@tempc \@tempb \let \@tempb \@tempa \fi \ifx
  \@tempb \@empty \def\@tempb {arXiv}\fi \@ifundefined
  {mn@eprint@\@tempb}{\@tempb:\@tempc}{\expandafter \expandafter \csname
  mn@eprint@\@tempb\endcsname \expandafter{\@tempc}}}

\bibitem[\protect\citeauthoryear{{Abazajian} et~al.,}{{Abazajian}
  et~al.}{2009}]{Abazajian09}
{Abazajian} K.~N.,  et~al., 2009, \mn@doi [\apjs]
  {10.1088/0067-0049/182/2/543}, \href
  {https://ui.adsabs.harvard.edu/abs/2009ApJS..182..543A} {182, 543}

\bibitem[\protect\citeauthoryear{{Allen}}{{Allen}}{1976}]{Allen76}
{Allen} C.~W.,  1976, {Astrophysical Quantities}

\bibitem[\protect\citeauthoryear{{Amblard} et~al.,}{{Amblard}
  et~al.}{2011}]{Amblard11}
{Amblard} A.,  et~al., 2011, \mn@doi [\nat] {10.1038/nature09771}, \href
  {https://ui.adsabs.harvard.edu/abs/2011Natur.470..510A} {470, 510}

\bibitem[\protect\citeauthoryear{{Ashby} et~al.,}{{Ashby}
  et~al.}{2009}]{Ashby09}
{Ashby} M.~L.~N.,  et~al., 2009, \mn@doi [\apj] {10.1088/0004-637X/701/1/428},
  \href {https://ui.adsabs.harvard.edu/abs/2009ApJ...701..428A} {701, 428}

\bibitem[\protect\citeauthoryear{{Audren}, {Lesgourgues}, {Benabed}  \&
  {Prunet}}{{Audren} et~al.}{2013}]{Audren13}
{Audren} B.,  {Lesgourgues} J.,  {Benabed} K.,   {Prunet} S.,  2013, \mn@doi
  [JCAP] {10.1088/1475-7516/2013/02/001}, \href
  {https://ui.adsabs.harvard.edu/abs/2013JCAP...02..001A} {2013, 001}

\bibitem[\protect\citeauthoryear{{Ben{\'\i}tez}}{{Ben{\'\i}tez}}{2000}]{Benitez00}
{Ben{\'\i}tez} N.,  2000, \mn@doi [\apj] {10.1086/308947}, \href
  {https://ui.adsabs.harvard.edu/abs/2000ApJ...536..571B} {536, 571}

\bibitem[\protect\citeauthoryear{{Bernstein}}{{Bernstein}}{2007}]{Bernstein07}
{Bernstein} R.~A.,  2007, \mn@doi [\apj] {10.1086/519824}, \href
  {https://ui.adsabs.harvard.edu/abs/2007ApJ...666..663B} {666, 663}

\bibitem[\protect\citeauthoryear{{Bernstein}, {Freedman}  \&
  {Madore}}{{Bernstein} et~al.}{2002a}]{Bernstein02a}
{Bernstein} R.~A.,  {Freedman} W.~L.,   {Madore} B.~F.,  2002a, \mn@doi [\apj]
  {10.1086/339422}, \href
  {https://ui.adsabs.harvard.edu/abs/2002ApJ...571...56B} {571, 56}

\bibitem[\protect\citeauthoryear{{Bernstein}, {Freedman}  \&
  {Madore}}{{Bernstein} et~al.}{2002b}]{Bernstein02b}
{Bernstein} R.~A.,  {Freedman} W.~L.,   {Madore} B.~F.,  2002b, \mn@doi [\apj]
  {10.1086/339423}, \href
  {https://ui.adsabs.harvard.edu/abs/2002ApJ...571...85B} {571, 85}

\bibitem[\protect\citeauthoryear{{Bernstein}, {Freedman}  \&
  {Madore}}{{Bernstein} et~al.}{2002c}]{Bernstein02c}
{Bernstein} R.~A.,  {Freedman} W.~L.,   {Madore} B.~F.,  2002c, \mn@doi [\apj]
  {10.1086/339424}, \href
  {https://ui.adsabs.harvard.edu/abs/2002ApJ...571..107B} {571, 107}

\bibitem[\protect\citeauthoryear{{Bernstein}, {Freedman}  \&
  {Madore}}{{Bernstein} et~al.}{2005}]{Bernstein05}
{Bernstein} R.~A.,  {Freedman} W.~L.,   {Madore} B.~F.,  2005, \mn@doi [\apj]
  {10.1086/444488}, \href
  {https://ui.adsabs.harvard.edu/abs/2005ApJ...632..713B} {632, 713}

\bibitem[\protect\citeauthoryear{{Bouchet}, {Lequeux}, {Maurice}, {Prevot}  \&
  {Prevot-Burnichon}}{{Bouchet} et~al.}{1985}]{Bouchet85}
{Bouchet} P.,  {Lequeux} J.,  {Maurice} E.,  {Prevot} L.,   {Prevot-Burnichon}
  M.~L.,  1985, \aap, \href
  {https://ui.adsabs.harvard.edu/abs/1985A&A...149..330B} {149, 330}

\bibitem[\protect\citeauthoryear{{Bullock}, {Kolatt}, {Sigad}, {Somerville},
  {Kravtsov}, {Klypin}, {Primack}  \& {Dekel}}{{Bullock}
  et~al.}{2001}]{Bullock01}
{Bullock} J.~S.,  {Kolatt} T.~S.,  {Sigad} Y.,  {Somerville} R.~S.,  {Kravtsov}
  A.~V.,  {Klypin} A.~A.,  {Primack} J.~R.,   {Dekel} A.,  2001, \mn@doi
  [\mnras] {10.1046/j.1365-8711.2001.04068.x}, \href
  {https://ui.adsabs.harvard.edu/abs/2001MNRAS.321..559B} {321, 559}

\bibitem[\protect\citeauthoryear{{Calzetti}, {Armus}, {Bohlin}, {Kinney},
  {Koornneef}  \& {Storchi-Bergmann}}{{Calzetti} et~al.}{2000}]{Calzetti00}
{Calzetti} D.,  {Armus} L.,  {Bohlin} R.~C.,  {Kinney} A.~L.,  {Koornneef} J.,
   {Storchi-Bergmann} T.,  2000, \mn@doi [\apj] {10.1086/308692}, \href
  {https://ui.adsabs.harvard.edu/abs/2000ApJ...533..682C} {533, 682}

\bibitem[\protect\citeauthoryear{{Cao} et~al.,}{{Cao} et~al.}{2018}]{Cao18}
{Cao} Y.,  et~al., 2018, \mn@doi [\mnras] {10.1093/mnras/sty1980}, \href
  {https://ui.adsabs.harvard.edu/abs/2018MNRAS.480.2178C} {480, 2178}

\bibitem[\protect\citeauthoryear{{Cao}, {Gong}, {Feng}, {Cooray}, {Cheng}  \&
  {Chen}}{{Cao} et~al.}{2020}]{Cao20}
{Cao} Y.,  {Gong} Y.,  {Feng} C.,  {Cooray} A.,  {Cheng} G.,   {Chen} X.,
  2020, \mn@doi [\apj] {10.3847/1538-4357/abada1}, \href
  {https://ui.adsabs.harvard.edu/abs/2020ApJ...901...34C} {901, 34}

\bibitem[\protect\citeauthoryear{{Cooray} et~al.,}{{Cooray}
  et~al.}{2012a}]{Cooray12}
{Cooray} A.,  et~al., 2012a, \mn@doi [\nat] {10.1038/nature11474}, \href
  {https://ui.adsabs.harvard.edu/abs/2012Natur.490..514C} {490, 514}

\bibitem[\protect\citeauthoryear{{Cooray}, {Gong}, {Smidt}  \&
  {Santos}}{{Cooray} et~al.}{2012b}]{Cooray12apj}
{Cooray} A.,  {Gong} Y.,  {Smidt} J.,   {Santos} M.~G.,  2012b, \mn@doi [\apj]
  {10.1088/0004-637X/756/1/92}, \href
  {https://ui.adsabs.harvard.edu/abs/2012ApJ...756...92C} {756, 92}

\bibitem[\protect\citeauthoryear{{Dube}, {Wickes}  \& {Wilkinson}}{{Dube}
  et~al.}{1977}]{Dube77}
{Dube} R.~R.,  {Wickes} W.~C.,   {Wilkinson} D.~T.,  1977, \mn@doi [\apjl]
  {10.1086/182475}, \href
  {https://ui.adsabs.harvard.edu/abs/1977ApJ...215L..51D} {215, L51}

\bibitem[\protect\citeauthoryear{{Dube}, {Wickes}  \& {Wilkinson}}{{Dube}
  et~al.}{1979}]{Dube79}
{Dube} R.~R.,  {Wickes} W.~C.,   {Wilkinson} D.~T.,  1979, \mn@doi [\apj]
  {10.1086/157292}, \href
  {https://ui.adsabs.harvard.edu/abs/1979ApJ...232..333D} {232, 333}

\bibitem[\protect\citeauthoryear{{Eisenhardt} et~al.,}{{Eisenhardt}
  et~al.}{2004}]{Eisenhardt04}
{Eisenhardt} P.~R.,  et~al., 2004, \mn@doi [\apjs] {10.1086/423180}, \href
  {https://ui.adsabs.harvard.edu/abs/2004ApJS..154...48E} {154, 48}

\bibitem[\protect\citeauthoryear{{Fernandez}, {Iliev}, {Komatsu}  \&
  {Shapiro}}{{Fernandez} et~al.}{2012}]{Fernandez12}
{Fernandez} E.~R.,  {Iliev} I.~T.,  {Komatsu} E.,   {Shapiro} P.~R.,  2012,
  \mn@doi [\apj] {10.1088/0004-637X/750/1/20}, \href
  {https://ui.adsabs.harvard.edu/abs/2012ApJ...750...20F} {750, 20}

\bibitem[\protect\citeauthoryear{{Fitzpatrick}}{{Fitzpatrick}}{1986}]{Fitzpatrick86}
{Fitzpatrick} E.~L.,  1986, \mn@doi [\aj] {10.1086/114237}, \href
  {https://ui.adsabs.harvard.edu/abs/1986AJ.....92.1068F} {92, 1068}

\bibitem[\protect\citeauthoryear{{Fixsen}, {Moseley}  \& {Arendt}}{{Fixsen}
  et~al.}{2000}]{Fixsen00}
{Fixsen} D.~J.,  {Moseley} S.~H.,   {Arendt} R.~G.,  2000, \mn@doi [\apjs]
  {10.1086/313390}, \href
  {https://ui.adsabs.harvard.edu/abs/2000ApJS..128..651F} {128, 651}

\bibitem[\protect\citeauthoryear{{Gong} et~al.,}{{Gong} et~al.}{2019}]{Gong19}
{Gong} Y.,  et~al., 2019, \mn@doi [\apj] {10.3847/1538-4357/ab391e}, \href
  {https://ui.adsabs.harvard.edu/abs/2019ApJ...883..203G} {883, 203}

\bibitem[\protect\citeauthoryear{{Grogin} et~al.,}{{Grogin}
  et~al.}{2011}]{Grogin11}
{Grogin} N.~A.,  et~al., 2011, \mn@doi [\apjs] {10.1088/0067-0049/197/2/35},
  \href {https://ui.adsabs.harvard.edu/abs/2011ApJS..197...35G} {197, 35}

\bibitem[\protect\citeauthoryear{{Helgason}, {Ricotti}  \&
  {Kashlinsky}}{{Helgason} et~al.}{2012}]{Helgason12}
{Helgason} K.,  {Ricotti} M.,   {Kashlinsky} A.,  2012, \mn@doi [\apj]
  {10.1088/0004-637X/752/2/113}, \href
  {https://ui.adsabs.harvard.edu/abs/2012ApJ...752..113H} {752, 113}

\bibitem[\protect\citeauthoryear{{Hivon}, {G{\'o}rski}, {Netterfield}, {Crill},
  {Prunet}  \& {Hansen}}{{Hivon} et~al.}{2002}]{Hivon02}
{Hivon} E.,  {G{\'o}rski} K.~M.,  {Netterfield} C.~B.,  {Crill} B.~P.,
  {Prunet} S.,   {Hansen} F.,  2002, \mn@doi [\apj] {10.1086/338126}, \href
  {https://ui.adsabs.harvard.edu/abs/2002ApJ...567....2H} {567, 2}

\bibitem[\protect\citeauthoryear{{Hoekstra}, {Viola}  \&
  {Herbonnet}}{{Hoekstra} et~al.}{2017}]{Hoekstra17}
{Hoekstra} H.,  {Viola} M.,   {Herbonnet} R.,  2017, \mn@doi [\mnras]
  {10.1093/mnras/stx724}, \href
  {https://ui.adsabs.harvard.edu/abs/2017MNRAS.468.3295H} {468, 3295}

\bibitem[\protect\citeauthoryear{{Koekemoer} et~al.,}{{Koekemoer}
  et~al.}{2011}]{Koekemoer11}
{Koekemoer} A.~M.,  et~al., 2011, \mn@doi [\apjs] {10.1088/0067-0049/197/2/36},
  \href {https://ui.adsabs.harvard.edu/abs/2011ApJS..197...36K} {197, 36}

\bibitem[\protect\citeauthoryear{{Krick} \& {Bernstein}}{{Krick} \&
  {Bernstein}}{2007}]{Krick07}
{Krick} J.~E.,  {Bernstein} R.~A.,  2007, \mn@doi [\aj] {10.1086/518787}, \href
  {https://ui.adsabs.harvard.edu/abs/2007AJ....134..466K} {134, 466}

\bibitem[\protect\citeauthoryear{{Lauer} et~al.,}{{Lauer}
  et~al.}{2021}]{Lauer21}
{Lauer} T.~R.,  et~al., 2021, \mn@doi [\apj] {10.3847/1538-4357/abc881}, \href
  {https://ui.adsabs.harvard.edu/abs/2021ApJ...906...77L} {906, 77}

\bibitem[\protect\citeauthoryear{{Lin}, {Mohr}  \& {Stanford}}{{Lin}
  et~al.}{2004}]{Lin04}
{Lin} Y.-T.,  {Mohr} J.~J.,   {Stanford} S.~A.,  2004, \mn@doi [\apj]
  {10.1086/421714}, \href
  {https://ui.adsabs.harvard.edu/abs/2004ApJ...610..745L} {610, 745}

\bibitem[\protect\citeauthoryear{{Madau}}{{Madau}}{1995}]{Madau95}
{Madau} P.,  1995, \mn@doi [\apj] {10.1086/175332}, \href
  {https://ui.adsabs.harvard.edu/abs/1995ApJ...441...18M} {441, 18}

\bibitem[\protect\citeauthoryear{{Matsuoka}, {Ienaka}, {Kawara}  \&
  {Oyabu}}{{Matsuoka} et~al.}{2011}]{Matsuoka11}
{Matsuoka} Y.,  {Ienaka} N.,  {Kawara} K.,   {Oyabu} S.,  2011, \mn@doi [\apj]
  {10.1088/0004-637X/736/2/119}, \href
  {https://ui.adsabs.harvard.edu/abs/2011ApJ...736..119M} {736, 119}

\bibitem[\protect\citeauthoryear{{Mattila}}{{Mattila}}{1976}]{Mattila76}
{Mattila} K.,  1976, \aap, \href
  {https://ui.adsabs.harvard.edu/abs/1976A&A....47...77M} {47, 77}

\bibitem[\protect\citeauthoryear{{Metropolis}, {Rosenbluth}, {Rosenbluth},
  {Teller}  \& {Teller}}{{Metropolis} et~al.}{1953}]{Metropolis53}
{Metropolis} N.,  {Rosenbluth} A.~W.,  {Rosenbluth} M.~N.,  {Teller} A.~H.,
  {Teller} E.,  1953, \mn@doi [\jcp] {10.1063/1.1699114}, \href
  {https://ui.adsabs.harvard.edu/abs/1953JChPh..21.1087M} {21, 1087}

\bibitem[\protect\citeauthoryear{{Miller} et~al.,}{{Miller}
  et~al.}{2013}]{Miller13}
{Miller} L.,  et~al., 2013, \mn@doi [\mnras] {10.1093/mnras/sts454}, \href
  {https://ui.adsabs.harvard.edu/abs/2013MNRAS.429.2858M} {429, 2858}

\bibitem[\protect\citeauthoryear{{Mitchell-Wynne} et~al.,}{{Mitchell-Wynne}
  et~al.}{2015}]{Mitchell-Wynne15}
{Mitchell-Wynne} K.,  et~al., 2015, \mn@doi [Nature Communications]
  {10.1038/ncomms8945}, \href
  {https://ui.adsabs.harvard.edu/abs/2015NatCo...6.7945M} {6, 7945}

\bibitem[\protect\citeauthoryear{{Planck Collaboration} et~al.,}{{Planck
  Collaboration} et~al.}{2016}]{Planck16XIII}
{Planck Collaboration} et~al., 2016, \mn@doi [\aap]
  {10.1051/0004-6361/201525830}, \href
  {https://ui.adsabs.harvard.edu/abs/2016A&A...594A..13P} {594, A13}

\bibitem[\protect\citeauthoryear{{Planck Collaboration} et~al.,}{{Planck
  Collaboration} et~al.}{2020}]{Planck18VI}
{Planck Collaboration} et~al., 2020, \mn@doi [\aap]
  {10.1051/0004-6361/201833910}, \href
  {https://ui.adsabs.harvard.edu/abs/2020A&A...641A...6P} {641, A6}

\bibitem[\protect\citeauthoryear{{Polletta} et~al.,}{{Polletta}
  et~al.}{2007}]{Polletta07}
{Polletta} M.,  et~al., 2007, \mn@doi [\apj] {10.1086/518113}, \href
  {https://ui.adsabs.harvard.edu/abs/2007ApJ...663...81P} {663, 81}

\bibitem[\protect\citeauthoryear{{Prevot}, {Lequeux}, {Maurice}, {Prevot}  \&
  {Rocca-Volmerange}}{{Prevot} et~al.}{1984}]{Prevot84}
{Prevot} M.~L.,  {Lequeux} J.,  {Maurice} E.,  {Prevot} L.,
  {Rocca-Volmerange} B.,  1984, \aap, \href
  {https://ui.adsabs.harvard.edu/abs/1984A&A...132..389P} {132, 389}

\bibitem[\protect\citeauthoryear{{Raichoor} et~al.,}{{Raichoor}
  et~al.}{2014}]{Raichoor14}
{Raichoor} A.,  et~al., 2014, \mn@doi [\apj] {10.1088/0004-637X/797/2/102},
  \href {https://ui.adsabs.harvard.edu/abs/2014ApJ...797..102R} {797, 102}

\bibitem[\protect\citeauthoryear{{Robin}, {Reyl{\'e}}, {Derri{\`e}re}  \&
  {Picaud}}{{Robin} et~al.}{2003}]{Robin03}
{Robin} A.~C.,  {Reyl{\'e}} C.,  {Derri{\`e}re} S.,   {Picaud} S.,  2003,
  \mn@doi [\aap] {10.1051/0004-6361:20031117}, \href
  {https://ui.adsabs.harvard.edu/abs/2003A&A...409..523R} {409, 523}

\bibitem[\protect\citeauthoryear{{Rowe} et~al.,}{{Rowe} et~al.}{2015}]{Rowe15}
{Rowe} B.~T.~P.,  et~al., 2015, \mn@doi [Astronomy and Computing]
  {10.1016/j.ascom.2015.02.002}, \href
  {https://ui.adsabs.harvard.edu/abs/2015A&C....10..121R} {10, 121}

\bibitem[\protect\citeauthoryear{{Seaton}}{{Seaton}}{1979}]{Seaton79}
{Seaton} M.~J.,  1979, \mn@doi [\mnras] {10.1093/mnras/187.1.73P}, \href
  {https://ui.adsabs.harvard.edu/abs/1979MNRAS.187P..73S} {187, 73}

\bibitem[\protect\citeauthoryear{{Simard} et~al.,}{{Simard}
  et~al.}{2002}]{Simard02}
{Simard} L.,  et~al., 2002, \mn@doi [\apjs] {10.1086/341399}, \href
  {https://ui.adsabs.harvard.edu/abs/2002ApJS..142....1S} {142, 1}

\bibitem[\protect\citeauthoryear{{Thacker} et~al.,}{{Thacker}
  et~al.}{2013}]{Thacker13}
{Thacker} C.,  et~al., 2013, \mn@doi [\apj] {10.1088/0004-637X/768/1/58}, \href
  {https://ui.adsabs.harvard.edu/abs/2013ApJ...768...58T} {768, 58}

\bibitem[\protect\citeauthoryear{{Thacker}, {Gong}, {Cooray}, {De Bernardis},
  {Smidt}  \& {Mitchell-Wynne}}{{Thacker} et~al.}{2015}]{Thacker15}
{Thacker} C.,  {Gong} Y.,  {Cooray} A.,  {De Bernardis} F.,  {Smidt} J.,
  {Mitchell-Wynne} K.,  2015, \mn@doi [\apj] {10.1088/0004-637X/811/2/125},
  \href {https://ui.adsabs.harvard.edu/abs/2015ApJ...811..125T} {811, 125}

\bibitem[\protect\citeauthoryear{{Ubeda} \& {et al.}}{{Ubeda} \& {et
  al.}}{2012}]{Ubeda12}
{Ubeda} L.,  {et al.} 2012, {Advanced Camera for Surveys Instrument Handbook
  for Cycle 21 v. 12.0}.
p.~12

\bibitem[\protect\citeauthoryear{{Viero} et~al.,}{{Viero}
  et~al.}{2013}]{Viero13}
{Viero} M.~P.,  et~al., 2013, \mn@doi [\apj] {10.1088/0004-637X/772/1/77},
  \href {https://ui.adsabs.harvard.edu/abs/2013ApJ...772...77V} {772, 77}

\bibitem[\protect\citeauthoryear{{Zemcov} et~al.,}{{Zemcov}
  et~al.}{2013}]{Zemcov13}
{Zemcov} M.,  et~al., 2013, \mn@doi [\apjs] {10.1088/0067-0049/207/2/31}, \href
  {https://ui.adsabs.harvard.edu/abs/2013ApJS..207...31Z} {207, 31}

\bibitem[\protect\citeauthoryear{{Zemcov} et~al.,}{{Zemcov}
  et~al.}{2014}]{Zemcov14}
{Zemcov} M.,  et~al., 2014, \mn@doi [Science] {10.1126/science.1258168}, \href
  {https://ui.adsabs.harvard.edu/abs/2014Sci...346..732Z} {346, 732}

\bibitem[\protect\citeauthoryear{{Zhan}}{{Zhan}}{2011}]{Zhan11}
{Zhan} H.,  2011, \mn@doi [Scientia Sinica Physica, Mechanica \& Astronomica]
  {10.1360/132011-961}, \href
  {https://ui.adsabs.harvard.edu/abs/2011SSPMA..41.1441Z} {41, 1441}

\bibitem[\protect\citeauthoryear{{Zhou} et~al.,}{{Zhou} et~al.}{2021}]{Zhou21}
{Zhou} X.,  et~al., 2021, \mn@doi [\apj] {10.3847/1538-4357/abda3e}, \href
  {https://ui.adsabs.harvard.edu/abs/2021ApJ...909...53Z} {909, 53}

\bibitem[\protect\citeauthoryear{{de Vaucouleurs}}{{de
  Vaucouleurs}}{1948}]{deVaucouleurs48}
{de Vaucouleurs} G.,  1948, Annales d'Astrophysique, \href
  {https://ui.adsabs.harvard.edu/abs/1948AnAp...11..247D} {11, 247}

\makeatother
\end{thebibliography}

%\begin{thebibliography}{99}
%\end{thebibliography}

%%%%%%%%%%%%%%%%%%%%%%%%%%%%%%%%%%%%%%%%%%%%%%%%%%

%%%%%%%%%%%%%%%%% APPENDICES %%%%%%%%%%%%%%%%%%%%%

%\appendix



%%%%%%%%%%%%%%%%%%%%%%%%%%%%%%%%%%%%%%%%%%%%%%%%%%

% Don't change these lines
\bsp	% typesetting comment
\label{lastpage}
\end{document}